# Approximation of Discs by Octagons on Pixel-Plane via Jaccard's Proximity Criterion: Theoretical Approach and Experimental Results Analysis

**Irakli Dochviri**
iraklidoch@yahoo.com
CAUCASUS INTERNATIONAL UNIVERSITY

**Alexander Gamkrelidze**
alexander.gamkrelidze@tsu.ge
TBILISI STATE UNIVERSITY

**Revaz Kurdiani**
revaz.kurdiani@tsu.ge
TBILISI STATE UNIVERSITY

2 November, 2022

# Table of Contents

**ABSTRACT:** In the present paper we study approximation of discs by octagons on the pixel plane. To decide which octagon approximates better the given disc we use Jaccard's distance. The table of Jaccard's distances (calculated by a software created for these purposes) are presented at the end of the paper. The results for proximity are given in the form of a graph. Some properties of considered octagons are also studied.

## 1. Introduction

Nowadays it is hard to imagine our life without electronic devices such as mobile phones, tablets, computers and other gadgets that have displays. The basic idea of how a display works is based on pixels. This is naturally related with digital geometry, where real shapes with continuous curves could be represented as discrete structures (see [Klette-Rosenfeld], [Chen], [Kiselman], [Gruenbaum-Shephard]). On the pixel plane to depict real geometric shapes we can use various mathematical approximations simultaneously with some logical assumptions. However, in computer graphics preference is given to the visual effects (in order to keep the perception of real shape). To show circular shapes on pixel displays one has to select in some sense best approximations, which depends on the criteria and features of the research tasks (see [McIliory], [Nakamura-Aizawa], [Andres], [Andres-Roussillon], [Pham]). In [Barrett] and [Sheppard] the authors consider a variety of pixelated shapes to approximate Euclidean discs with small diameters. We have a general approach to this problem and consider algorithmically constructable pixelated discs. Main purpose of the present paper is an approximation of the Euclidean disc on the pixel plane with certain types of polygons. We introduce a new approach for approximation of shapes on the pixel plane via the Jaccard's distance (see [Jaccard]) minimality criterion. In particular, we calculate Jaccard's distances between a given disc and thus octagons that are treated as approximations of this disc. Then, we choose the minimal one from these distances and the octagon which corresponds to the chosen distance is declared as the best approximation for this disc. Note that, in [Peters-Kordzaya-Dochviri] Jaccard's distance was used to study intersections of analytical pixel discs in the $l_1$ metrics.

Among various forms of pixels (square, rectangular, triangular, hexagonal, etc.) the most commonly used in practice are square pixels. However, pixels of other forms can provide better approximation for some shapes. Also, triangular and hexagonal pixels allow us to use some complicated geometric properties of shapes in discrete cases. So, various authors consider and study not only square pixels ([Tirunelveli-Gordon-Pistorius], [Nagy]). In the present article, we consider square pixels, but the same problem can be formulated for other types of pixels. This may be the subject of further work.

As we mentioned above, we try to approximate discs with a certain type of polygon. More precisely, we consider a family of octagons and calculate Jaccard's distance between these octagons and pixelated discs. The octagon, which is closest to a given disc, is the best approximation for that disc. In general, we get that the octagon which is the best approximation for a given disc has 95.5% intersection with it (regarding the area of the union). So, we can say that, we approximate the discs with 95.5% proximity. In future, we are going to approximate the discs with other polygons or shapes and observe the difference in proximity. This is a new task and generates the ideas for new articles.

In Section 2 we give some well-known notions and facts and in Section 3 we describe pixelated discs.

In section 4 we introduce a family of octagons for approximation of pixelated discs. These octagons in the paper are called disc-like octagons. We give some examples of disc-like octagons and describe their structure.

In Section 5 we provide explicit formulas for perimeters and areas of disc-like octagons. Also, we describe the relationship between perimeter and area of disc-like octagons.

In Section 6 we discuss how we conducted the experiments and calculations to estimate proximity of disc-like octagons and discs. At the end of the section we analyze obtained results and present some interesting cases in more detail.

Section 7 contains the table of Jaccard's distances between disc-like octagons and discs and in Section 8 the graph describing proximity is presented.

## 2. Preliminaries

In this section we are going to give some well-known definitions that we need throughout the paper.

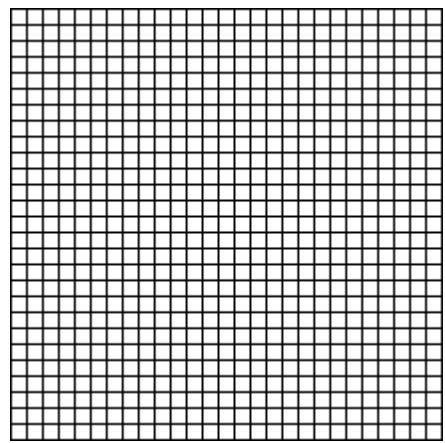
Picture 1.

Our main aim is to analyze and investigate shapes drawn with square pixels. To formalize our study, we use the mathematical model of the pixel plane, which is a two dimensional plane divided into infinitely many equal squares like the chessboard is divided into 64 black and white squares. In other words, let us draw "horizontal" and "vertical" parallel lines on the Euclidean plane, so that there is the same equal distance between the adjacent lines. In this way, we get the splitting of the plane into the squares obtained by intersecting these lines. A finite piece of this plane is shown on Picture 1.

We call each square a pixel. A pixel can have its own color and the whole pixel must have the same color.

To give a more precise mathematical definition of the pixel plane we can use the digital plane.

**Definition 1.** Let us consider the Euclidean plane $R^2$ with the usual distance. The digital plane is subspace $Z^2$ of the Euclidean plane.

Around the point $(i, j)$ in the digital plane one can draw the square with vertices

$$\left(i - \tfrac{1}{2}, j - \tfrac{1}{2}\right), \left(i + \tfrac{1}{2}, j - \tfrac{1}{2}\right), \left(i + \tfrac{1}{2}, j + \tfrac{1}{2}\right), \left(i - \tfrac{1}{2}, j + \tfrac{1}{2}\right).$$

In this way, the digital plane can be viewed as the pixel plane. The point of the digital plane can be identified with the pixel (i.e the square) around it and vice versa the pixels can be identified with their centers. One of the advantages of this point of view is that the pixel plane gets an induced metric from the digital plane. In this metric pixels have unit length.

The distance between two pixels is calculated as the distance between their centers. The distance from a pixel to a point of the Euclidean plane is calculated as distance from the center of the pixel to that point.

Two pixels are said to be neighbors if they share a common side (horizontal or vertical). It is clear that each pixel has 4 neighbor pixels (see Picture 2).

Picture 2. Black colored pixels are the neighbors to the gray colored pixel.

In terms of digital planes, two points $(i, j)$ and $(i', j')$ are neighbors if and only if $|i' - i| + |j' - j| = 1$.

**Definition 1.** A collection of pixels is called a pixelated shape.

**Definition 1.**

1) A pixel of a pixelated shape is called a bounding-line pixel if it has at least one neighbor that is not a pixel of the same pixelated shape.
2) The collection of bounding-line pixels is called the bounding-line of the pixelated shape.
3) The quantity of bounding-line pixels is called the perimeter of the pixelated shape.
4) The quantity of all pixels of pixelated shape is called the area of the pixelated shape.

To illustrate pixelated shapes on the pixel plane we use different colors. Thus pixels that belong to one particular shape can be painted in one or more colors. For example, we can use different colors for the bounding-line of a shape and for the shape itself. In the case of several shapes being presented in the same picture, we use different colors to distinguish each shape and each intersection of these shapes.

The introduced notions are demonstrated in the following picture

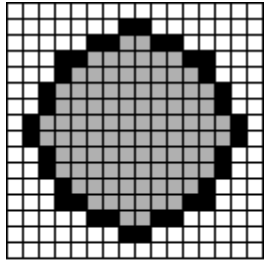

d = 14
P = 36
A = 132

Picture 3. The shape consists of all black and gray pixels and the black pixels are the boundary-line pixels. Perimeter equals 36 and the area is equal to 132.

We say that an Euclidean square (an ordinary square on the Euclidean plane) is a *proper square* on the pixel plane if the vertices of the square coincide to some vertices of pixels. In other words, a proper square is union of $n^2$ (n by n) pixels. The quantity of pixels along the side of a proper square is called the length of the side of the proper square.

For a given pixelated shape a *covering square* is a proper square which contains all pixels of the pixelated shape. A covering square of the given pixelated shape, with minimal length of side, is called an *enveloping square* of the pixelated shape. The length of the side of an enveloping square is called the *diameter* of the pixelated shape.

Let us consider an ordinary line L on the Euclidean plane. A pixelated line associated with L is the collection of thus pixels that contain some segment of the line L.

We will use the phrases Euclidean shape or Euclidean line to indicate that we mean a shape or line on the continuous Euclidean plane rather than a pixelated shape or line. One has to distinguish between pixelated and Euclidean shapes or lines. For example, an Euclidean line means just an ordinary line on the Euclidean plane, while a pixelated line is a collection of pixels.

One can convert any Euclidean shape to a pixelated shape in various ways. We have chosen the following approach. Let F be an Euclidean shape. The pixelated shape F' obtained from Euclidean shape F is the collection of thus pixels whose centers are in F.

Using this approach we can define pixelated convex polygons and pixelated discs.

Let n be an integer and $L_i$ be Euclidean lines for each $i = 1,...,n$. Let us choose a semi-plane $H_i$ for each line $L_i$. An Euclidean convex polygon is the intersection of the semi-planes $H_i$.

**Definition 1.** A pixelated convex polygon is a pixelated shape obtained from an Euclidean convex polygon.

We are interested in investigating pixelated convex octagons to approximate pixelated discs with those octagons. Pixelated octagons are studied in sections 4 and 5.

**Definition 1.** Let us assume that X is the center (respectively a vertex) of a pixel and d is an odd (respectively even) integer. The disk on the pixel plane (or pixelated disk) with the center X and diameter d is the collection of such pixels P that the distance from the center of P to X is less or equal to d/2.

Recall that Jaccard's distance J between finite sets A and B is defined as $J(A,B) = 1 - \frac{\#(A \cap B)}{\#(A \cup B)}$, where #(A∩B) and #(A∪B) are the number of elements in the intersection and union of A and B, respectively.

One can consider a pixelated shape as a set of pixels and apply Jaccard's distance to pixelated shapes.

It is known that the floor (respectively ceiling) of a real number n is denoted as $\lfloor n \rfloor$ (respectively $\lceil n \rceil$) and is defined by the formula $\lfloor n \rfloor = max\{m \in Z \mid m \leq n\}$ (respectively $\lceil n \rceil = min\{m \in Z \mid n \leq m\}$).

## 3. Images of Pixelated Discs

In this section we want to give some examples of discs on the pixel plane. On each picture, a disc consists of all black and gray pixels while black pixels are the pixels of the boundary-line of the disc. At the bottom of pictures the diameter of the disc is indicated. We start with pictures of discs with diameters from 1 to 15.

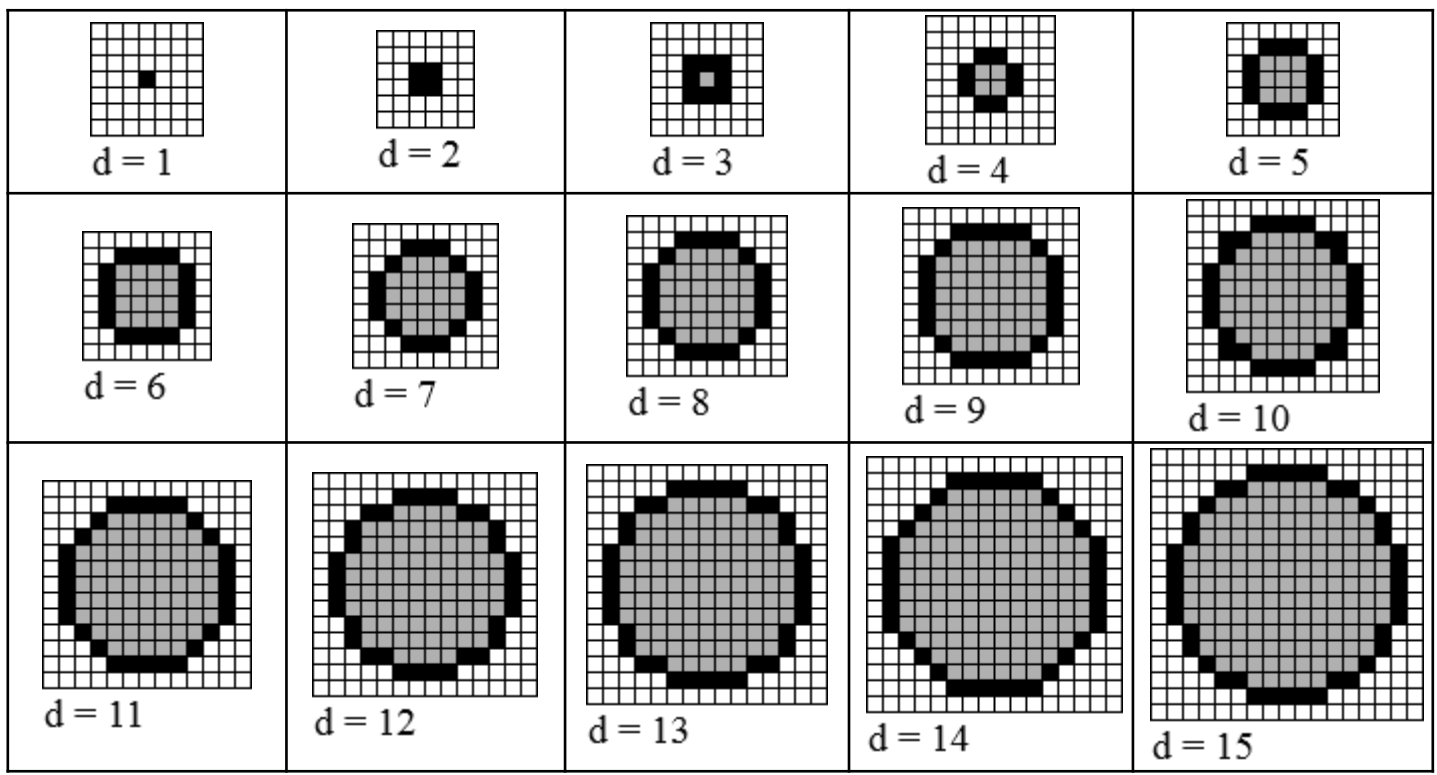

Some other discs with larger diameters ($d = 50, 55, 99, 100$) are given below. The diameters again are indicated at the bottom of the pictures

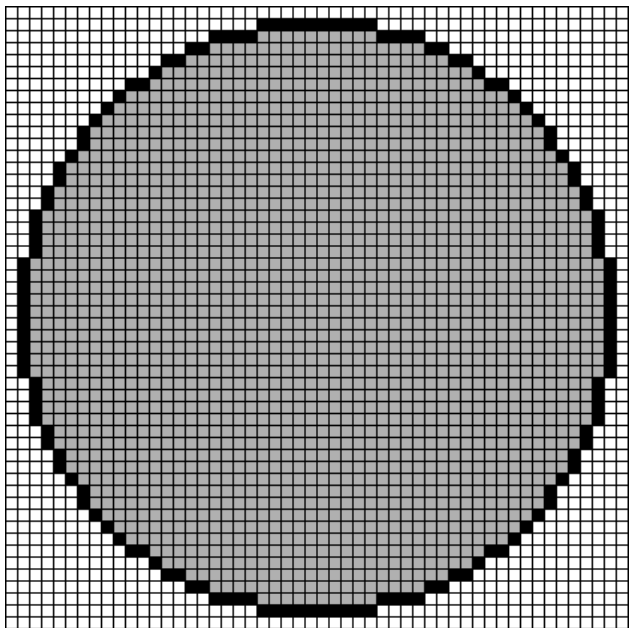
d = 50

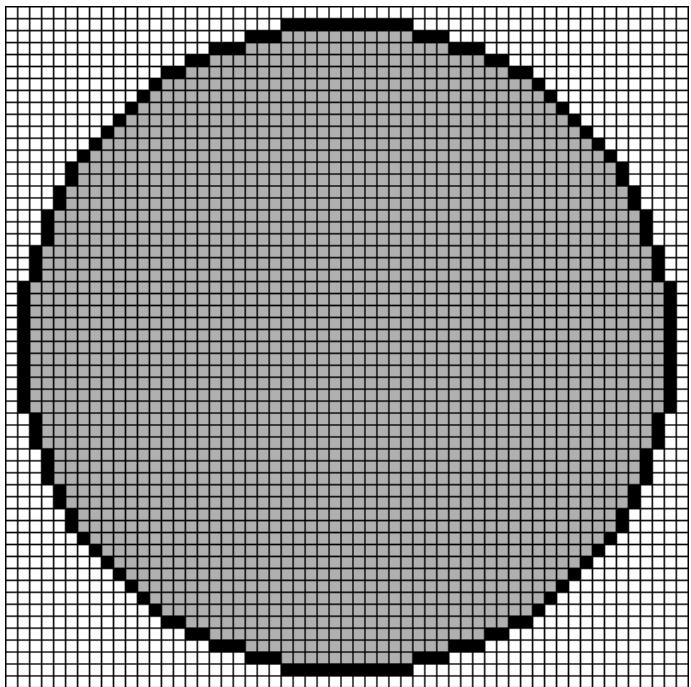
d = 55

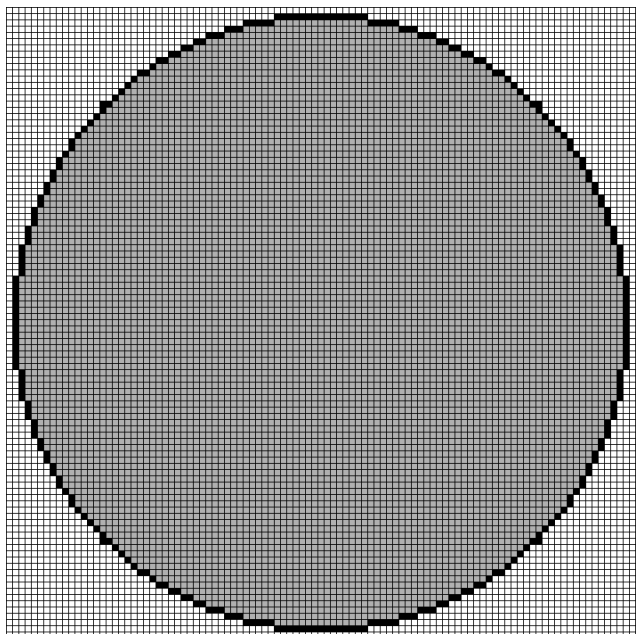
d = 99

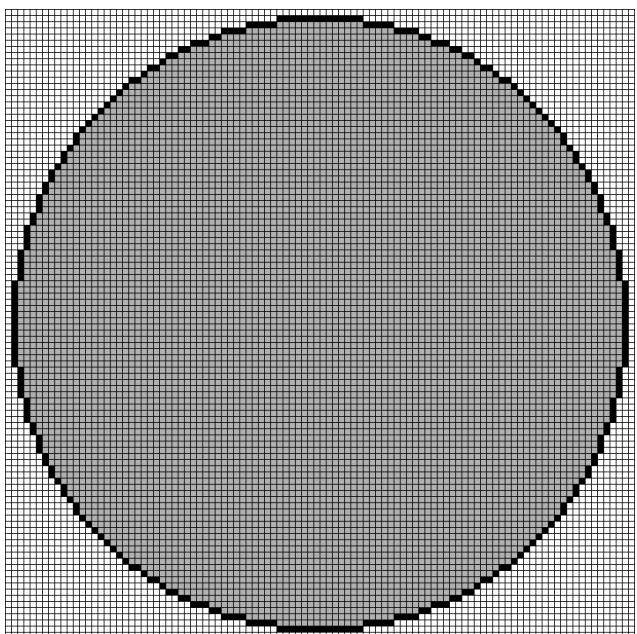
d = 100

## 4. Pixelated Octagons for Approximation of Discs

In this section we introduce a family of pixelated octagons. These octagons are used in next sections for approximation of pixelated discs.

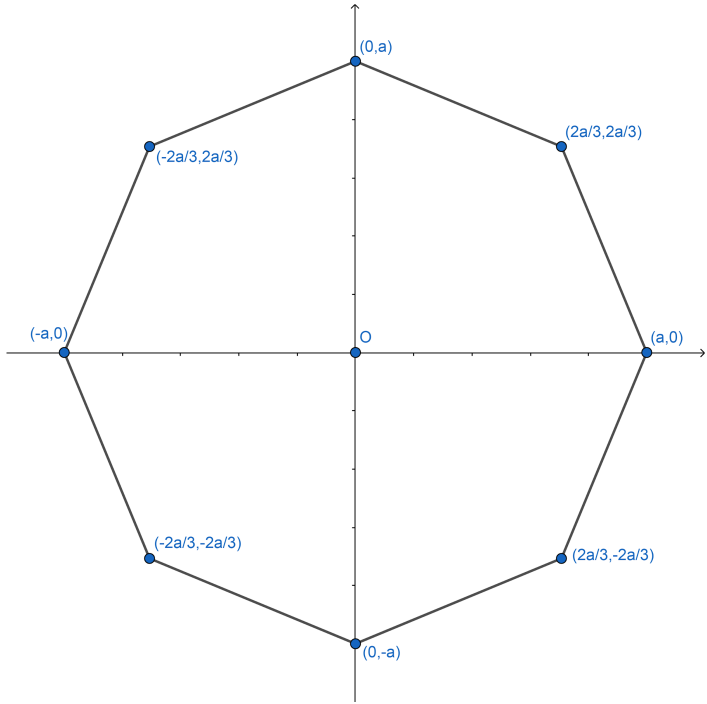

Picture 4.

Let us consider a family of Euclidean octagons depending on the parameter $a \in R$. The vertices of an octagon of this family are given by the coordinates on the Cartesian plane as shown on the picture 4.

These octagons are equilateral but irregular. All eight sides of an octagon of such type are equal to each other. But, angles are different. The octagons are radially and centrally symmetrical. The origin O is the center of symmetry of the octagons and is called the center of the octagon. Four vertices are located on the axes and the remaining four vertices are on the bisectors of the quarters of the Cartesian plane.

Depending on the position of this octagon on the pixel plane we will get different pixelated octagons. If we put such an octagon on the pixel plane in the way that axes are parallel to the sides of pixels and the origin O coincides with the center of a pixel, then we get a pixelated octagon with odd diameter. Alternatively, if we put such an octagon on the pixel plane in the way that axes are parallel to the sides of pixels and the origin O coincides with a vertex of a pixel, then we get a pixelated octagon with even diameter.

Let $d$ be a positive integer. We can choose $a = \frac{d-0.5}{2}$ to get the pixelated octagon with diameter $d$. Depending on parity of $d$, we have to adjust the origin O to the center or to a vertex of a pixel. One of the pixelated octagons obtained in this way is demonstrated in Picture 5.

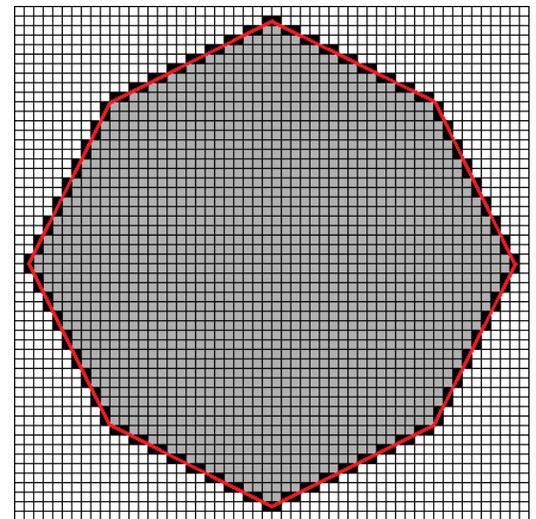

d = 52
P = 136
A = 1804

Picture 5.

More examples of these octagons are shown in the images below, where $d$ is the diameter, $P$ is the pixelated perimeter, and $A$ is the pixelated area of the octagon. The octagon's boundary-line is shown in black color and the inner pixels in gray.

| 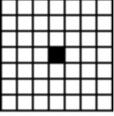 d = 1 P = 1 A = 1 | 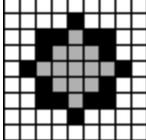 d = 7 P = 16 A = 29 | 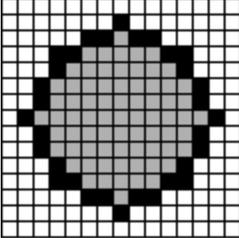 d = 13 P = 32 A = 105 | 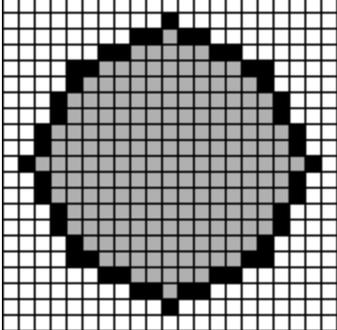 d = 19 P = 48 A = 229 |
|---|---|---|---|
| 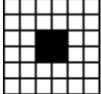 d = 2 P = 4 A = 4 | 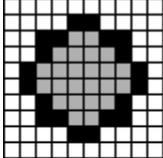 d = 8 P = 20 A = 44 | 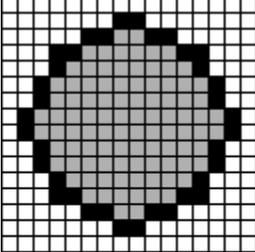 d = 14 P = 36 A = 132 | 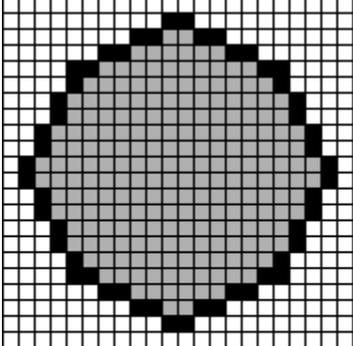 d = 20 P = 52 A = 268 |
| 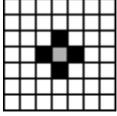 d = 3 P = 4 A = 5 | 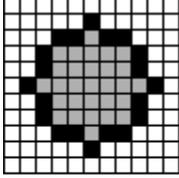 d = 9 P = 20 A = 49 | 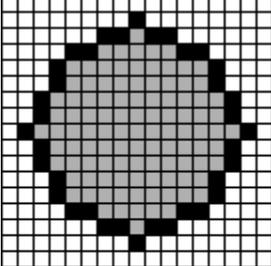 d = 15 P = 36 A = 141 | 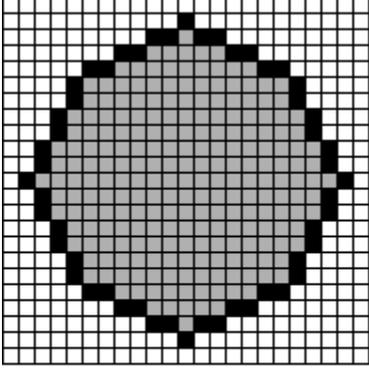 d = 21 P = 52 A = 281 |

| 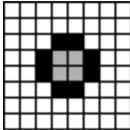 d = 4 P = 8 A = 12 | 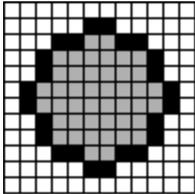 d = 10 P = 24 A = 68 | 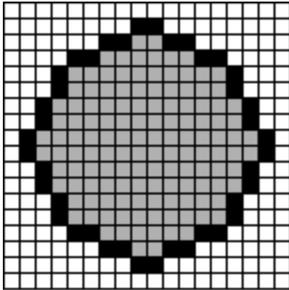 d = 16 P = 40 A = 172 | 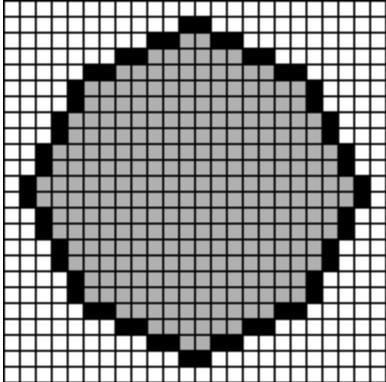 d = 22 P = 56 A = 324 |
|---|---|---|---|
| 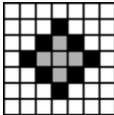 d = 5 P = 8 A = 13 | 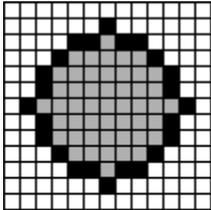 d = 11 P = 24 A = 73 | 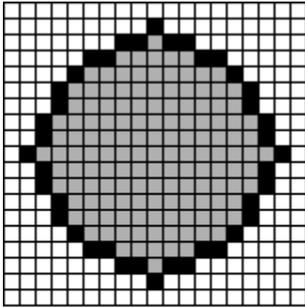 d = 17 P = 40 A = 181 | 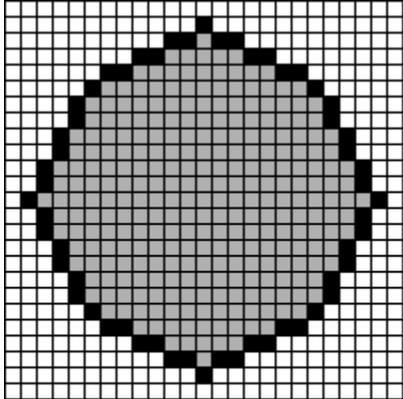 d = 23 P = 56 A = 337 |
| 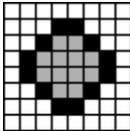 d = 6 P = 12 A = 24 | 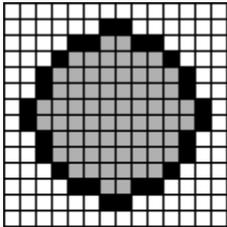 d = 12 P = 28 A = 96 | 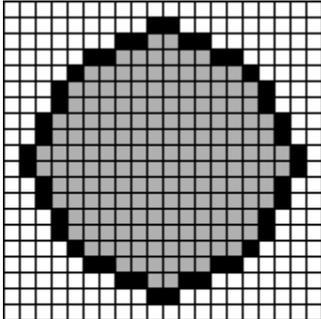 d = 18 P = 44 A = 216 | 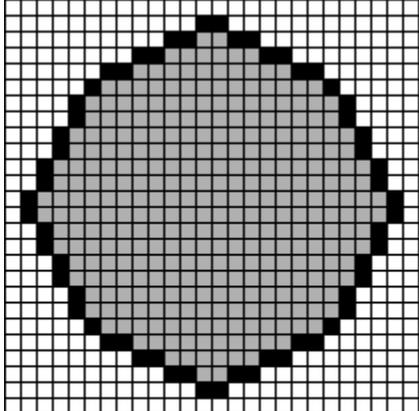 d = 24 P = 60 A = 384 |

One can observe that the sides of the pixelated octagons are formed with the following pixelated lines

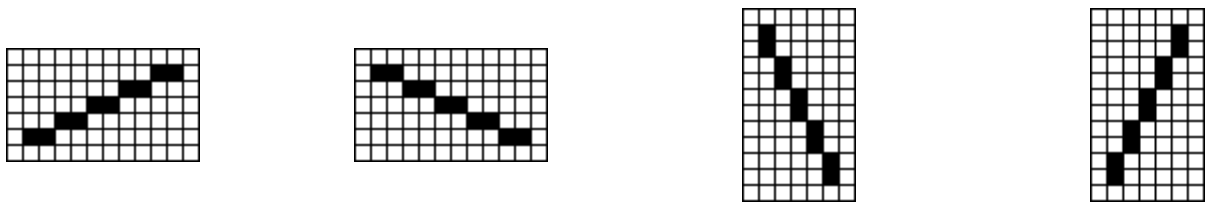

These lines consist of blocks of two pixels oriented either horizontally or vertically. For a given octagon, the number of these blocks in each line are the same and this number is a characteristic of the octagon.

Other characteristics of the given octagon are types of vertices. The vertices on the axes have two types shown below

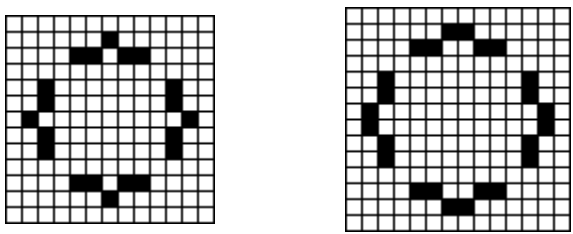

The vertices on bisectors of quarters of the Cartesian plane have three types shown below

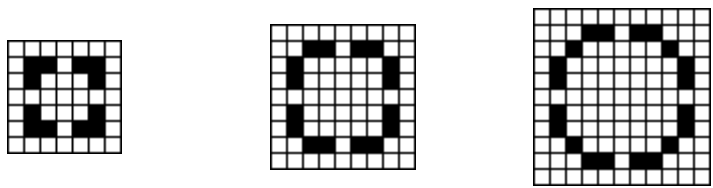

Pairing these types together we get 6 types. Two octagons have the same type iff the difference between the diameters is divisible by 6 and one can classify the octagons by the residue of diameter modulo 6. Let us denote by $d$ the diameter of the given octagon. In the following table we give the description of types of our octagons.

| Diameter | Type of octagon | Type of the vertices on the axes | Type of the vertices on bisectors |
|---|---|---|---|
| $d = 6n + 1$ | 1 | 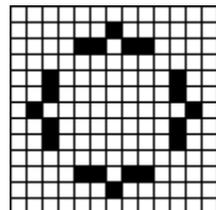 | 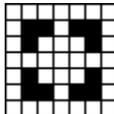 |
| $d = 6n + 2$ | 2 | 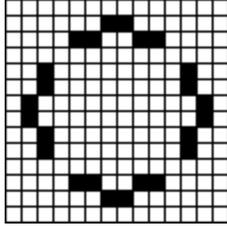 | 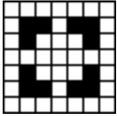 |
| $d = 6n + 3$ | 3 | 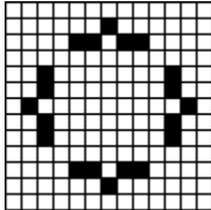 | 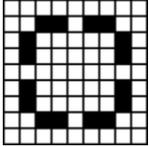 |
| $d = 6n + 4$ | 4 | 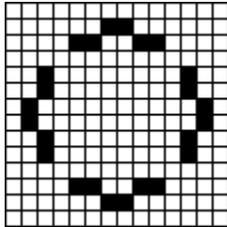 | 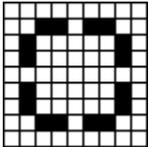 |
| $d = 6n + 5$ | 5 | 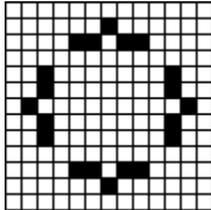 | 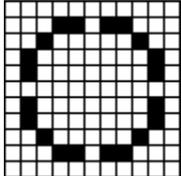 |
| $d = 6n$ | 0 | 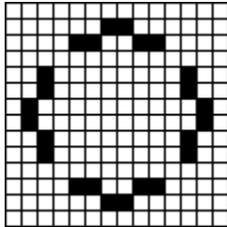 | 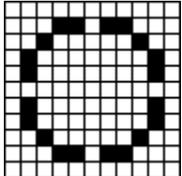 |

In the consecutive sections we consider discs and octagons with the same centers on the digital plane and have interest in investigating their proximity from the viewpoint of Jaccard's distance estimation. Below are presented octagonal approximations of discs. On the pictures $o$ is the diameter of the octagon and $c$ is the diameter of the circle, while $j$ is the Jaccard's distance between these two shapes.

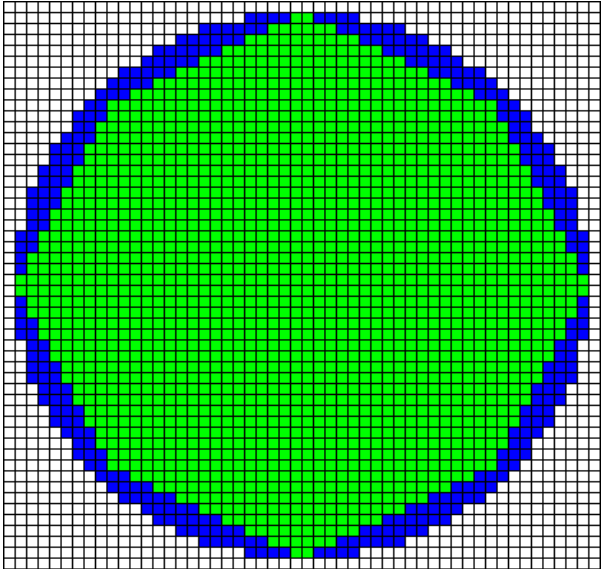
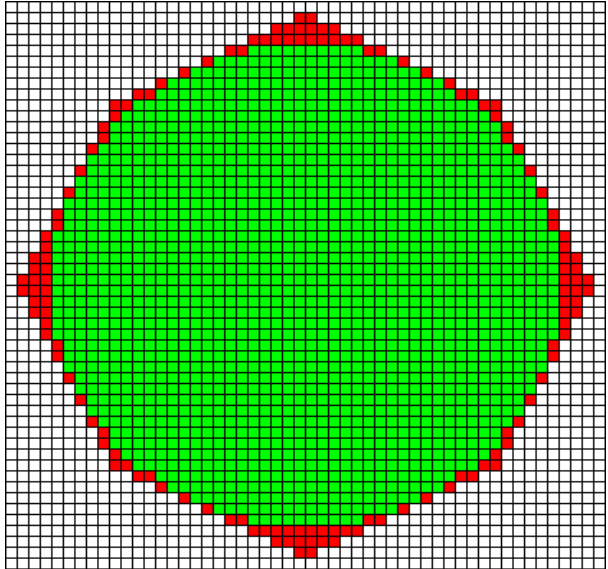
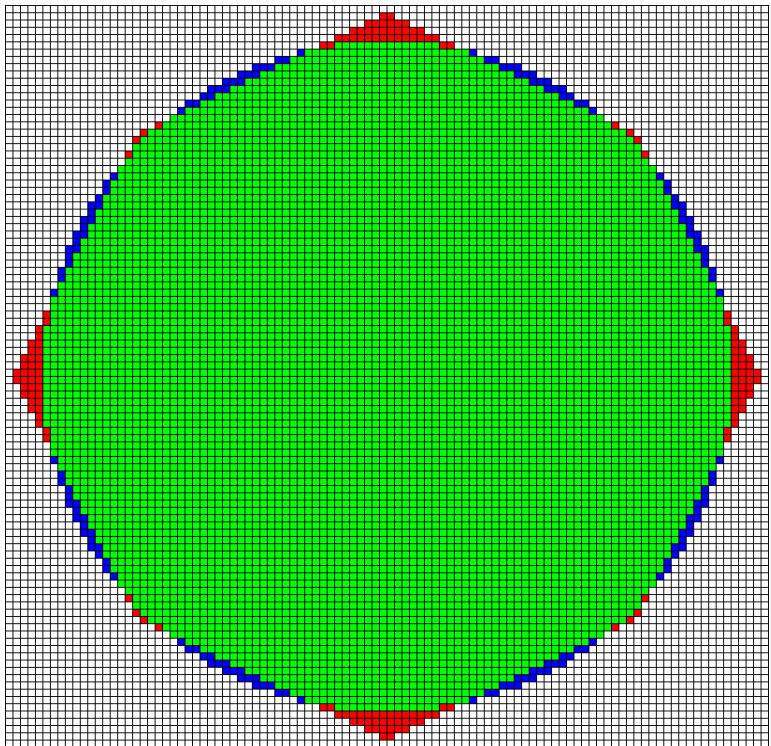

Picture 6. Green area is the intersection of two shapes, blue pixels are the pixels of the disc not belonging to the octagon and red pixels are the pixels of the octagon not belonging to the disc.

As we mentioned above, the pixelated octagons discussed in this section will be used for approximation of pixelated discs. For this reason, in the rest of the article we call them disc-like octagons.

## 5. Properties of Disc-like Octagons.

In this section we investigate and describe some properties of disc-like octagons.

The following result establishes the relationship between the perimeter and the diameter of a disc-like octagon.

**Theorem 1.** Let $d > 1$ be a diameter of a disc-like octagon, then perimeter can be calculated via the formula

$$P(d) = 2d + 4\lfloor \tfrac{d}{6} \rfloor - 2, \text{ if } d \text{ is odd}$$

$$= 2d + 4\lceil \tfrac{d}{6} \rceil - 4, \text{ if } d \text{ is even}$$

*Remark:* If d is even then the formula $P(d) = 2d + 4\lceil \tfrac{d}{6} \rceil - 4$ can be rewritten as follows

$$P(d) = 2d + 4\lfloor \tfrac{d}{6} \rfloor, \text{ if } d \text{ is even and not divisible by 6}$$

$$= 2d + 4\lfloor \tfrac{d}{6} \rfloor - 4, \text{ if } d \text{ is divisible by 6,}$$

or compactly could be written as

$$P(d) = 5d - 14\lfloor \tfrac{d}{6} \rfloor - 4 - \tfrac{1}{2}\left(d - 6\lfloor \tfrac{d}{6} \rfloor\right)^2.$$

**Proof.** Let us consider following six cases of $d = 6n + i$, where $i = 1, 3, 5, 0, -2, -4$ it is obvious that these cover all possibilities.

If $i = 1, 3, 5$ then $d$ is odd and we shall prove $P(d) = 2d + 4\lfloor \tfrac{d}{6} \rfloor - 2$. For $d = 6n + 1$ the formula above can be rewritten as

$$P(6n + 1) = 2(6n + 1) + 4\lfloor \tfrac{6n+1}{6} \rfloor - 2 = 12n + 4n = 16n.$$

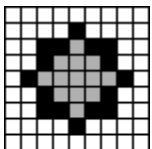

d = 7
P = 16
A = 29

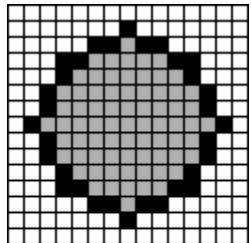

d = 13
P = 32
A = 105

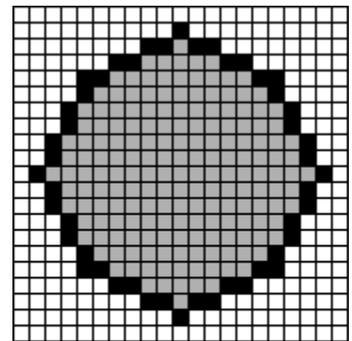

d = 19
P = 48
A = 229

When $n = 1, 2, 3$ we get $d = 7, 13, 19$ respectively and corresponding disc-like octagons are shown above.

Observe that $n$ is the number of two pixel blocks on the side of a disc-like octagon. We will use induction on $n$.

Base case: $n = 1$ and $d = 7$. Computing by formula we get $P(7) = 16 \cdot 1 = 16$, which is true.

Now, assume that, the formula is true for $n$ and prove it for $n + 1$. Observe that, when $n$ increases by one, then each side of the disc-like octagon gets an additional two pixel block. So, the perimeter will increase by 8 times 2 pixels, which is 16. Formalizing this, we get

$$P(6(n+1)+1) = P(6n+1) + 16 = 16n + 16 = 16(n+1),$$

which proves the case. Other cases can be proved similarly and are left to the reader as an exercise.

**Corollary 1.** The perimeter of disk-like octagon with diameter $d$ can be written alternatively as

$$P(d) = \tfrac{8}{3}d - 4, \ \ \textit{if d is divisible by 6}$$

$$= \tfrac{8}{3}d - \tfrac{4}{3}, \ \ \textit{if } (d-2) \textit{ is divisible by 6}$$

$$= \tfrac{8}{3}d - \tfrac{8}{3}, \ \ \textit{if } (d-4) \textit{ is divisible by 6}$$

$$= \tfrac{8}{3}d - \tfrac{8}{3}, \ \ \textit{if } (d-1) \textit{ is divisible by 6}$$

$$= \tfrac{8}{3}d - 4, \ \ \textit{if } (d-3) \textit{ is divisible by 6}$$

$$= \tfrac{8}{3}d - \tfrac{16}{3}, \ \ \textit{if } (d-5) \textit{ is divisible by 6}$$

**Proof:** Directly follows from the theorem.

**Theorem 2.** Let $d$ be a diameter of a disc-like octagon then

$$A(d) = \tfrac{d^2}{2} + d - 6\lfloor\tfrac{d}{6}\rfloor^2 - 6\lfloor\tfrac{d}{6}\rfloor + 2d\lfloor\tfrac{d}{6}\rfloor, \ \ \textit{if d is even}$$

$$= \tfrac{d^2}{2} - 6\lfloor\tfrac{d}{6}\rfloor^2 - 4\lfloor\tfrac{d}{6}\rfloor + 2d\lfloor\tfrac{d}{6}\rfloor + \tfrac{1}{2}, \ \ \textit{if d is odd}$$

**Remark.** Observe that if $d$ is even one can rewrite the formula as follows

$$A(d) = \tfrac{d^2}{2} - d - 6\lceil\tfrac{d}{6}\rceil^2 + 6\lceil\tfrac{d}{6}\rceil + 2d\lceil\tfrac{d}{6}\rceil$$

Also, simplified view of above written formula is given bellow

$$A(d) = \tfrac{d^2}{2} + d + 2\lfloor\tfrac{d}{6}\rfloor\left(d - 3 - 3\lfloor\tfrac{d}{6}\rfloor\right), \ \ \textit{if d is even}$$

$$= \tfrac{d^2}{2} + \tfrac{1}{2} + 2\lfloor\tfrac{d}{6}\rfloor\left(d - 2 - 3\lfloor\tfrac{d}{6}\rfloor\right), \ \ \textit{if d is odd}$$

**Proof.** As in the case of perimeter, we consider separately six cases of diameter divisibility by 6. Let $d = 6n + i$, where $i = 0, 1,..., 5$. We prove the case of $i = 1$. Other cases can be proved similarly and are left to the reader.

One can rewrite the formula in the following form

$$A(d) = \tfrac{2}{3}d^2 - \tfrac{2}{3}d + 1, \text{ when } d = 6n + 1.$$

Let us use induction on $n$ as in the case of perimeter.

Base case: $n = 0$ and $d = 1$, then by formula $A(1) = 1$, which is true.

Now, assume that, the formula is true for $n$ and prove it for $n + 1$. For better understanding let us look at the picture 7, which presents two concentric disc-like octagons with diameters 13 and 19 pixels.

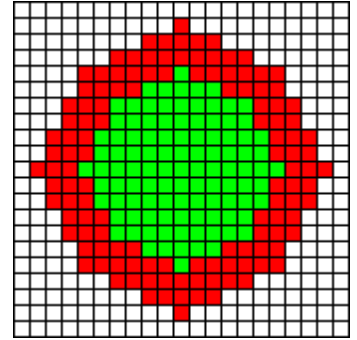

Picture 7.

The octagon with diameter 13 consists of the green pixels and the octagon with diameter 19 consists of all green and red pixels. So, we can calculate the area of the larger octagon by adding the area of the smaller octagon with the number of red pixels. Let us repaint the picture to count red pixels. We get a new picture (see Picture 8), where some of the red pixels are recolored blue and aqua.

Regardless the size of $n$ there will be 4 blue colored blocks of 3 pixels, 4 aqua colored blocks of 4 pixels and 8 times $2n$ red colored blocks of 3 pixels. Summing up all together we get

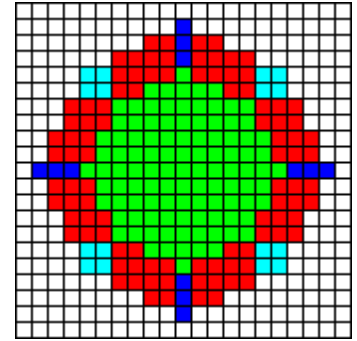

$$A(d + 6) = 4 \cdot 3 + 4 \cdot 4 + 8 \cdot 2n \cdot 3 + A(d) = 28 + 48n + A(d).$$

Picture 8.

On the other hand

$$A(d+6) - A(d) = \left(\tfrac{2}{3}(d+6)^2 - \tfrac{2}{3}(d+6) + 1\right) - \left(\tfrac{2}{3}d^2 - \tfrac{2}{3}d + 1\right)$$

$$= \tfrac{2}{3}(6(2d+6) - 6)$$

$$= 4((2d+6) - 1)$$

$$= 4(2d + 5)$$

$$= 4(2(6n+1) + 5)$$

$$= 4(12n + 7)$$

$$= 48n + 28.$$

This proves the case.

**Corollary 2.** The area of disc-like octagon with diameter $d$ can be written alternatively as

$$A(d) = \tfrac{2}{3}d^2, \text{ if } d \text{ is divisible by } 6$$

$$= \tfrac{2}{3}d^2 + \tfrac{4}{3}, \text{ if } (d-2) \text{ is divisible by } 6$$

$$= \tfrac{2}{3}d^2 + \tfrac{4}{3}, \text{ if } (d-4) \text{ is divisible by } 6$$

$$= \tfrac{2}{3}d^2 - \tfrac{2}{3}d + 1, \text{ if } (d-1) \text{ is divisible by } 6$$

$$= \tfrac{2}{3}d^2 - \tfrac{2}{3}d + 1, \text{ if } (d-3) \text{ is divisible by } 6$$

$$= \tfrac{2}{3}d^2 - \tfrac{2}{3}d - \tfrac{1}{3}, \text{ if } (d-5) \text{ is divisible by } 6$$

**Proof:** Directly follows from the theorem.

In the Euclidean plane, we observe that the ratio of the length of the circle to the perimeter of the encircled square is equal to the ratio of area of the corresponding disc to the area of the same square. This ratio is constant $\tfrac{\pi}{4} = \tfrac{2\pi r}{8r} = \tfrac{\pi R^2}{(2R)^2}$. In some sense, similar result holds for the disc-like octagons:

**Theorem 3.** Let $A(d)$ and $P(d)$ be the area and the perimeter respectively, of a disc-like octagon with diameter $d$, then

$$\lim_{d \to \infty} \frac{A(d)}{d^2} = \lim_{d \to \infty} \frac{P(d)}{4d} = \frac{2}{3}.$$

**Proof.** Immediately follows from Corollary 1 and Corollary 2.

## 6. Proximity Estimation via Computer Experiments

In this section we analyze proximities between disc-like octagons and discs. We wish to approximate the discs by these octagons. Under approximation of the disc, we mean the process of choosing among the disc-like octagons of various diameters the nearest one (regarding Jaccard's distance) to the given disc. For these reasons we made and conducted experiments on the computer to calculate Jaccard's distance between the pixelated discs and disc-like octagons. These experiments are carried out via the software which we developed especially for these purposes. Currently, we are adding a user-friendly interface to the program and we intend to publish it as a website for practical use. In the experiment we take two pixelated shapes: synchronized disc-like octagon and disc. Here synchronization means that the octagon and disc have the same (concentric) center. To evaluate Jaccard's distance between these shapes we calculate pixelated areas of their intersection $A(i)$ and union $A(u)$. Then Jaccard's distance $J$ is given by the formula

$$J = 1 - \frac{A(i)}{A(u)}.$$

The shapes are determined by the diameters and to choose the pair of shapes we consider pairs of integers $(d_c, d_o)$, where $d_c$ is the diameter of a pixelated disc and $d_o$ is the diameter of a disc-like octagon.

Let us notice that if one of the shapes is a subset of another then it does not make sense to increase the biggest one or decrease the smallest one, since in these cases the Jaccard's distance will be increased and therefore we get an uninteresting case. In the case of diameters are equal, synchronized disc contains the disc-like octagon and we can use the following restriction $d_c \leq d_o$, or otherwise $\frac{d_c}{d_o} \leq 1$.

To establish lower bound for the ratio $\frac{d_c}{d_o}$ we can observe the case when the disc is a subset of the disc-like octagon. For this reason, let

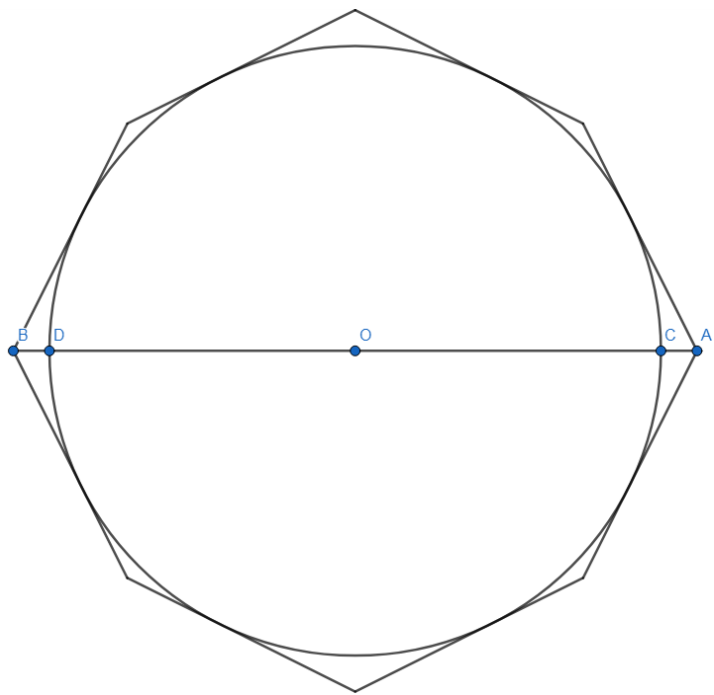

Picture 9.

us find the largest disc which is a subset of a fixed disc-like octagon and calculate the ratio of the diameters. The case of the Euclidean plane is described in Picture 9.

Here $AB$ is the diameter of the octagon and $CD$ is the diameter of the disc. One can show with simple calculations that the ratio $\frac{CD}{AB} = \frac{2}{\sqrt{5}} \approx 0.8944271909999$. However, in the pixelated cases this ratio can vary. If diameters are relatively small the ratio can be very different from the value of the Euclidean case. For example, if the disc-like octagon diameter $d_o$ is 21, or between 23 and 38 then the diameter of (maximal) subscribed disc $d_c = d_o - 4$. Calculating some of ratios in these cases we get the following values: $\frac{17}{21} = 0.(809523)$ and $\frac{19}{23} = 0.(8260869565217391304347)$.

The other values for a diameter of disc-like octagon $d_o$ and the diameter of (maximal) subscribed disc $d_c$ are given here

$$d_o = 39, \qquad d_c = 33$$
$$d_o = 40, \qquad d_c = 36$$
$$d_o = 41, \qquad d_c = 35$$
$$d_o = 42, \qquad d_c = 38$$
$$d_c = d_o - 6 \qquad if\ 43 \leq d_o \leq 60\ except\ d_o = 57\ and\ 59.$$

On the basis of general considerations when the number of pixels increases the ratio $\frac{d_c}{d_o}$ (here $d_o$ is the diameter of a disc-like octagon and $d_c$ is the diameter of the maximal subscribed disc in the octagon) gets closer to the above established number $\frac{2}{\sqrt{5}}$. Note that for $d_o = 50$ the diameter of the maximal subscribed disc is $d_c = 44$ and the ratio $\frac{d_c}{d_o} = 0.88$, which is very close to the value $\frac{2}{\sqrt{5}}$. Alternatively, an interesting example is the following: when the diameter of the disc-like octagon is 57, then the diameter of the maximal subscribed disc is 49. In this case, the ratio is $\frac{d_c}{d_o} = 0.(859649122807017543)$ and the difference from the value $\frac{2}{\sqrt{5}}$ is a little bit bigger than in the previous case. Hence, we established appropriate bounds for the ratio of diameters

$$k \leq \frac{d_c}{d_o} \leq 1,$$

where $k$ can be choose equal to 0.83 when $d_o$ is enough large ($d_o > 41$).
Alternatively, above restrictions can be viewed as $kd_o \leq d_c \leq d_o$. For small values of diameters, we can make another choice of restrictions

$$d_o - a \leq d_c \leq d_o,$$

where it is sufficient to choose $a = 7$, when $d_o \leq 41$.

Summarizing above stated, we perform calculations for the pairs of diameters $(d_o, d_c)$ in the following cases

$$1 \leq d_o \leq 41,\ max\{1, d_o - a\} \leq d_c \leq d_o \text{ and}$$
$$41 < d_o < 250,\ kd_o \leq d_c \leq d_o,$$

where $a$ and $k$ are the numbers given above.

If the difference between the diameters of synchronized disc-like octagon and disc is an odd number, then the centers of the shapes cannot be concentric as illustrated in Picture 10 and Picture 11. In the picture 10 are shown two shapes: the disc-like octagon with diameter 24, which consists of red and green pixels, while the disc with diameter 23 consists of blue and green pixels. In Picture 11 we have presented the disc-like octagon with diameter 45, which consists of red and green pixels, while the disc with diameter 38 consists of green pixels.

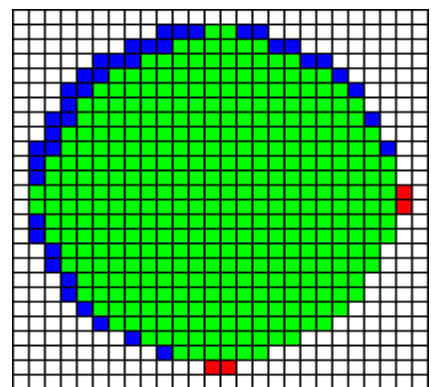

o = 24
c = 23
j = 0.10588235294118

Picture 10.

For this reason, we consider pairs of diameters $(d_o, d_c)$, where $d_o - d_c$ is an even number. This guarantees that we can choose the same center for both shapes. So, in our experiment we choose a pair of integers $(d_o, d_c)$ subject to the above mentioned restrictions.
Then consider a synchronized disc-like octagon and disc with the diameters corresponding to this pair. Finally, the Jaccard's distance between these two shapes is calculated. The experiments were conducted for all pairs satisfying the restrictions and the results were assembled into a proximity matrix. Part of this matrix is presented at the end of the article in section 7.

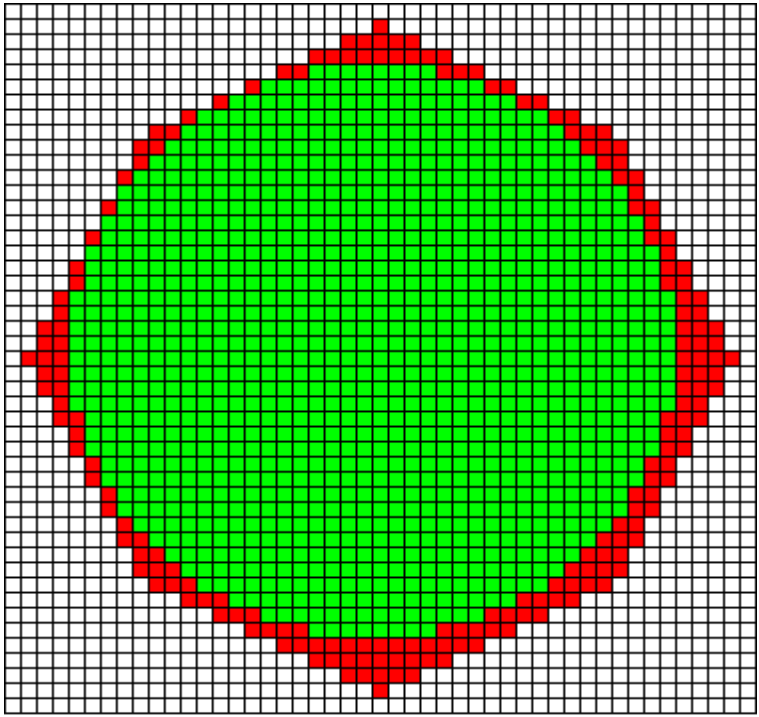

**Remark**: When the diameters take the following values

$$d_o = d_c = 1$$

$$d_o = d_c = 2$$

$$d_o = d_c = 4$$

then the shapes – the disc–like octagon and the disc are exactly the same. Consequently, Jaccard's distances in these exceptional cases are 0. In other cases, the Jaccard's distance is positive.

In the following table we present some interesting values of Jaccard's distances

o = 45
c = 38
j = 0.14912944738834

Picture 11.

| Diameters of octagons | Diameters of discs | Jaccard's distance |
|---:|---:|---:|
| 15 | 13 | 0.028 |
| 17 | 15 | 0.022 |
| 18 | 16 | 0.037 |
| 19 | 17 | 0.017 |
| 22 | 20 | 0.024 |
| 37 | 33 | 0.031 |
| 41 | 37 | 0.036 |
| 48 | 44 | 0.035 |

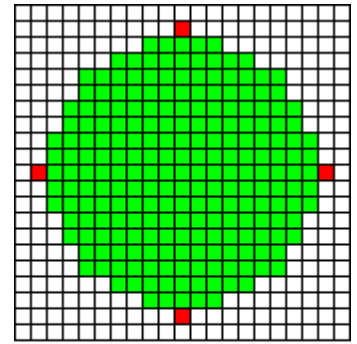

o = 19
c = 17
j = 0.0174672489082

Picture 12.

These values are smaller than others. Usually, other values are above 0.04. The smallest positive value is achieved when

$$d_o = 19 \text{ and } d_c = 17.$$

Picture 12 illustrates this case.

We analyze the obtained data in the following way:

- For each disc we are looking for the disc-like octagon which is nearest to the given disc regarding Jaccard's distance. This is achieved by choosing the minimal value in the appropriate column of the proximity matrix.
- For each disc-like octagon we are looking for the disc which is nearest to the given octagon regarding Jaccard's distance. This is achieved by choosing the minimal value in the appropriate row of the proximity matrix.

These calculations are performed by another software, which finds minimal values in each column and row of the proximity matrix.

We use graphs to illustrate the result in an effective way and it is presented at the end of the article, in Section 8.

Mainly, there are pairs of disc-like octagon and disc which are nearest to each other. More precisely, in most cases, when we take the nearest disc-like octagon for a given disc and then take the nearest disc for this octagon, we get the initial disc. Starting from a disc-like octagon and choosing the nearest disc for it, then taking the nearest disc-like octagon for this disc, we get the initial disc-like octagon again in the majority of cases. Nevertheless, in some cases, we can observe the opposite situation: starting from a disc and choosing the nearest disc-like octagon, then taking the nearest disc for this octagon, we can get a different disc from the given one. Also, if we start from a disc-like octagon and take the nearest disc for it, then take the nearest disc-like octagon to this disc, the obtained octagon can be different from the given one. These kinds of situations appear not very often and here we illustrate one example of such cases. More examples can be observed on the proximity graph.

Let us look at the piece of proximity graph

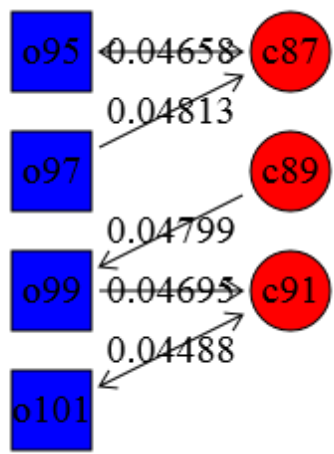

Picture 13.

The pictures below demonstrate that to the disc with diameter 89 nearest is the disc-like octagon with diameter 99.

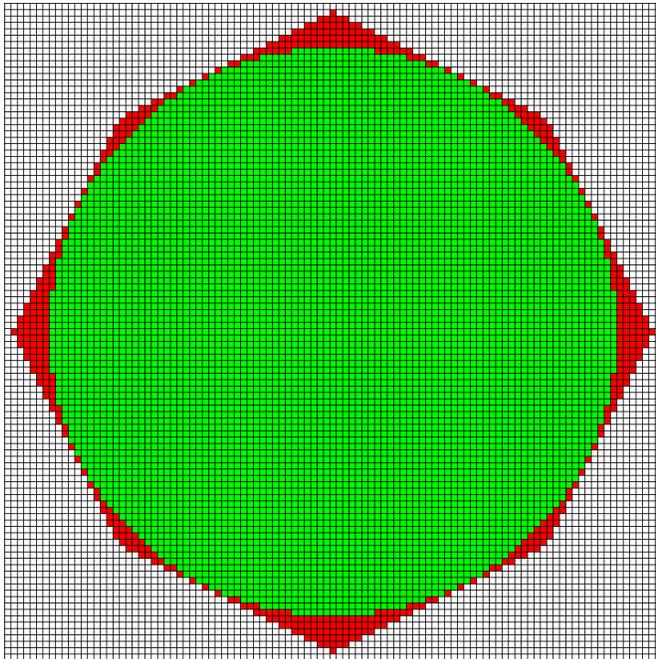
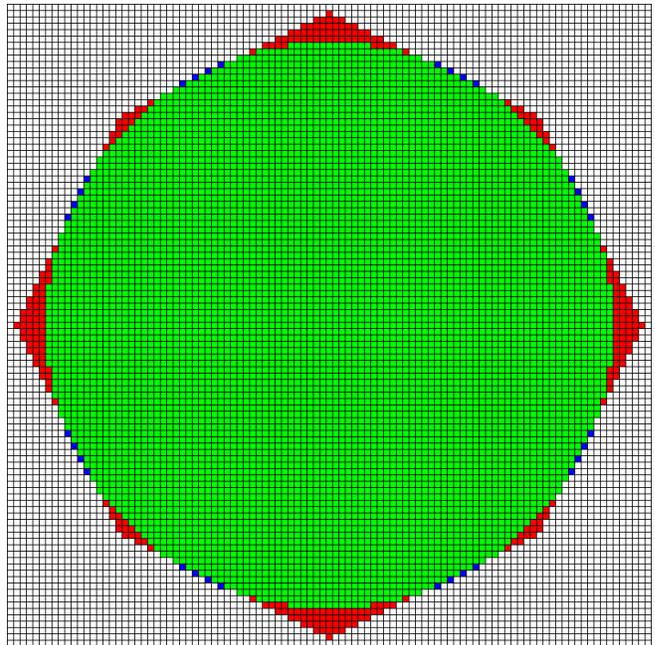
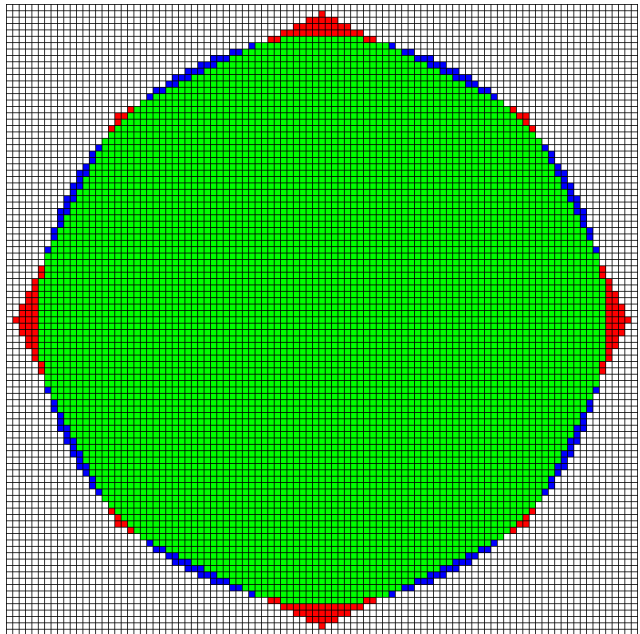

Picture 14.

Now let us explore the disc-like octagon with diameter 99 and find the nearest disc, we get the disc with diameter 91. This is illustrated in the following pictures

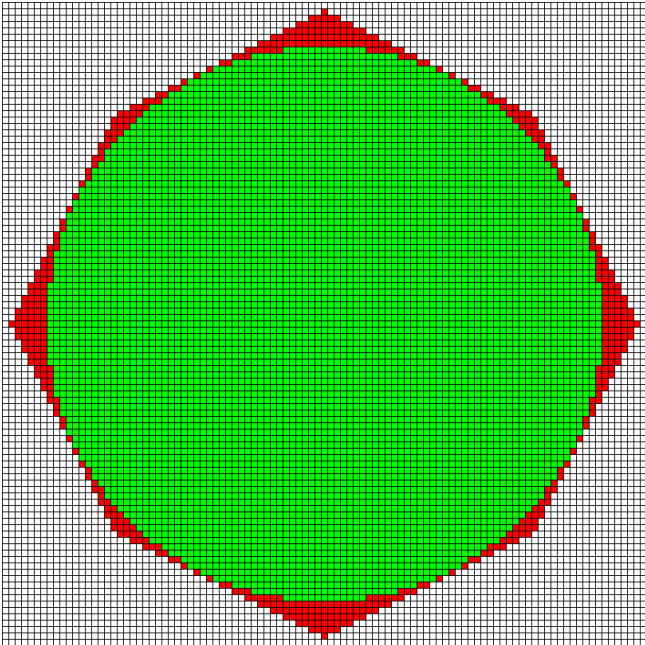

o = 99
c = 87
j = 0.079146699644458

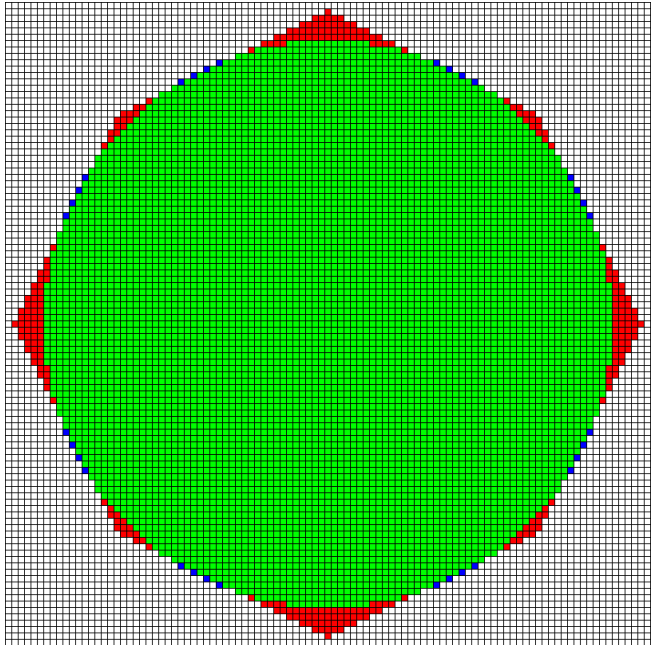

o = 99
c = 89
j = 0.047992616520535

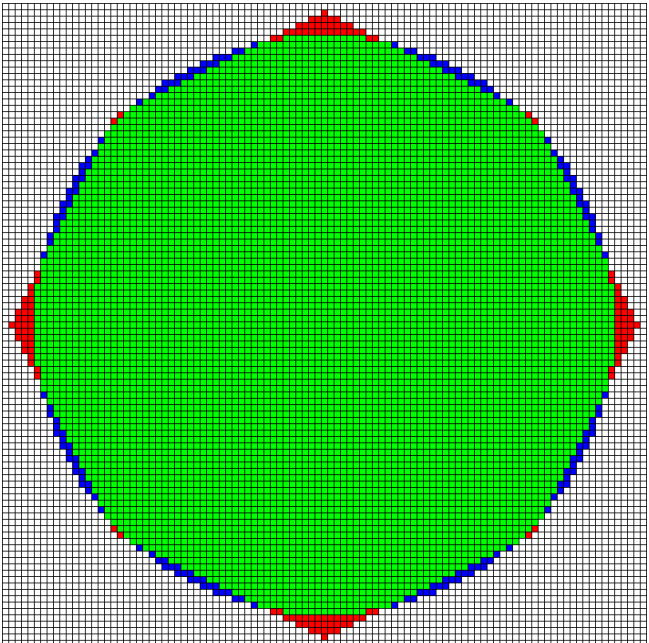

o = 99
c = 91
j = 0.046952595936795

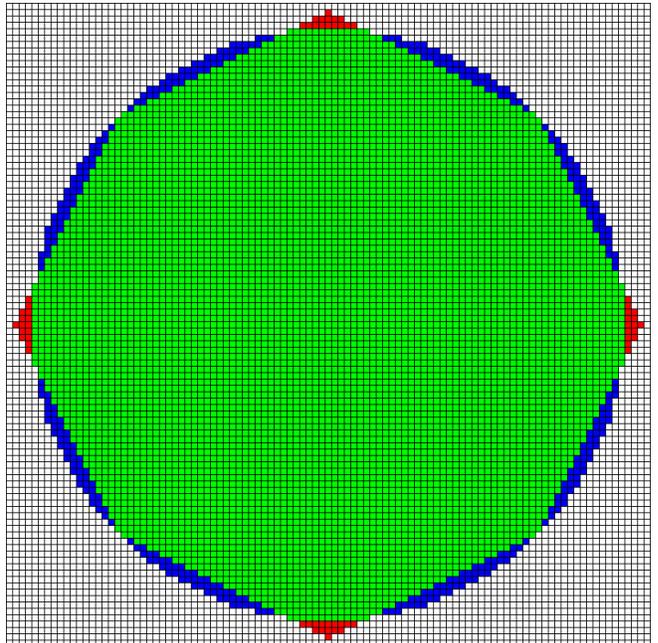

o = 99
c = 93
j = 0.063696128560993

Picture 15.

# 7. Proximity matrix

In this section we present values of Jaccard's distance between the disc-like octagons and discs. On vertical are listed diameters of disc-like octagons and on horizontal are diameters of discs.

| 0 | 1 | 2 | 3 | 4 | 5 | 6 | 7 | 8 | 9 | 10 | 11 | 12 | 13 | 14 | 15 | 16 | 17 | 18 | 19 | 20 |
|---|---|---|---|---|---|---|---|---|---|---|---|---|---|---|---|---|---|---|---|---|
| 1 | 0 | | | | | | | | | | | | | | | | | | | |
| 2 | | 0 | | | | | | | | | | | | | | | | | | | |
| 3 | 0,8 | | 0,444444 | | | | | | | | | | | | | | | | | | |
| 4 | | 0,666667 | | 0 | | | | | | | | | | | | | | | | | |
| 5 | 0,923077 | | 0,307692 | | 0,380952 | | | | | | | | | | | | | | | | |
| 6 | | 0,833333 | | 0,5 | | 0,25 | | | | | | | | | | | | | | | |
| 7 | 0,965517 | | 0,689655 | | 0,275862 | | 0,216216 | | | | | | | | | | | | | | |
| 8 | | 0,909091 | | 0,727273 | | 0,272727 | | 0,153846 | | | | | | | | | | | | | |
| 9 | | | 0,816327 | | 0,571429 | | 0,244898 | | 0,289855 | | | | | | | | | | | | |
| 10 | | | | 0,823529 | | 0,529412 | | 0,235294 | | 0,15 | | | | | | | | | | | |
| 11 | | | | | 0,712329 | | 0,493151 | | 0,054795 | | 0,247423 | | | | | | | | | | |
| 12 | | | | | | 0,666667 | | 0,458333 | | 0,166667 | | 0,142857 | | | | | | | | | |
| 13 | | | | | | | 0,647619 | | 0,342857 | | 0,076191 | | 0,233577 | | | | | | | | |
| 14 | | | | | | | 0,606061 | | 0,393939 | | 0,151515 | | 0,028369 | | 0,153846 | | | | | | |
| 15 | | | | | | | | 0,510638 | | 0,312057 | | 0,028369 | | 0,20339 | | | | | | | |
| 16 | | | | | | | | | 0,534884 | | 0,348837 | | 0,093023 | | 0,173077 | | | | | | |
| 17 | | | | | | | | | | 0,464088 | | 0,243094 | | 0,022099 | | 0,195556 | | | | | |
| 18 | | | | | | | | | | | 0,481482 | | 0,277778 | | 0,037037 | | 0,15625 | | | | |
| 19 | | | | | | | | | | | | 0,401747 | | 0,227074 | | 0,017467 | | 0,21843 | | | |
| 20 | | | | | | | | | | | | | 0,41791 | | 0,223881 | | 0,044776 | | | 0,151899 | |
| 21 | | | | | | | | | | | | | | 0,370107 | | 0,199288 | | 0,06734 | | | |
| 22 | | | | | | | | | | | | | | | 0,358025 | | 0,209877 | | | 0,024691 | |
| 23 | | | | | | | | | | | | | | | | 0,332344 | | 0,130564 | | | |
| 24 | | | | | | | | | | | | | | | | | 0,333333 | | | 0,177083 | |
| 25 | | | | | | | | | | | | | | | | | | 0,269327 | | | |
| 26 | | | | | | | | | | | | | | | | | | | 0,300885 | | |

Table 1.

| 0 | 21 | 22 | 23 | 24 | 25 | 26 | 27 | 28 | 29 | 30 | 31 | 32 | 33 | 34 | 35 | 36 | 37 | 38 | 39 | 40 |
|---|---|---|---|---|---|---|---|---|---|---|---|---|---|---|---|---|---|---|---|---|
| 21 | 0,194842 | | | | | | | | | | | | | | | | | | | |
| 22 | | 0,15625 | | | | | | | | | | | | | | | | | | | |
| 23 | 0,056657 | | 0,199525 | | | | | | | | | | | | | | | | | | |
| 24 | | 0,040816 | | 0,142857 | | | | | | | | | | | | | | | | | |
| 25 | 0,129676 | | 0,065882 | | 0,179959 | | | | | | | | | | | | | | | | |
| 26 | | 0,150443 | | 0,043478 | | 0,162963 | | | | | | | | | | | | | | | |
| 27 | 0,255864 | | 0,102345 | | 0,056795 | | 0,187175 | | | | | | | | | | | | | | |
| 28 | | 0,267176 | | 0,145038 | | 0,058394 | | 0,149351 | | | | | | | | | | | | | |
| 29 | | | 0,221812 | | 0,096118 | | 0,075732 | | 0,186466 | | | | | | | | | | | | |
| 30 | | | | 0,253333 | | 0,1 | | 0,051282 | | 0,162011 | | | | | | | | | | | |
| 31 | | | | | 0,21256 | | 0,070854 | | 0,077728 | | 0,170895 | | | | | | | | | | |
| 32 | | | | | | 0,210526 | | 0,099415 | | 0,066298 | | 0,157636 | | | | | | | | | |
| 33 | | | | | | | 0,18156 | | 0,056738 | | 0,069057 | | 0,181185 | | | | | | | | |
| 34 | | | | | | | | 0,202073 | | 0,072539 | | 0,068293 | | 0,153509 | | | | | | | |
| 35 | | | | | | | | 0,161412 | | 0,055486 | | 0,087861 | | 0,184995 | | | | | | | |
| 36 | | | | | | | | | 0,171296 | | 0,060185 | | 0,069565 | | 0,152941 | | | | | | |
| 37 | | | | | | | | | 0,15748 | | 0,031496 | | 0,094166 | | 0,180645 | | | | | | |
| 38 | | | | | | | | | | 0,157676 | | 0,053942 | | 0,070039 | | 0,142349 | | | | | |
| 39 | | | | | | | | | | | 0,129424 | | 0,047761 | | 0,095501 | | 0,17652 | | | | |
| 40 | | | | | | | | | | | | 0,146067 | | 0,044944 | | 0,063604 | | 0,155063 | | | |
| 41 | | | | | | | | | | | | | 0,10979 | | 0,036069 | | 0,096266 | | | | |
| 42 | | | | | | | | | | | | | | 0,132653 | | 0,044218 | | 0,081761 | | | |
| 43 | | | | | | | | | | | | | | | 0,099585 | | 0,042311 | | | | |
| 44 | | | | | | | | | | | | | | | | 0,130031 | | 0,045872 | | | |
| 45 | | | | | | | | | | | | | | | | | 0,09084 | | | | |
| 46 | | | | | | | | | | | | | | | | | | 0,104816 | | | |
| 47 | | | | | | | | | | | | | | | | | | | | | |
| 48 | | | | | | | | | | | | | | | | | | | 0,177083 | | |

Table 2.

| 0 | 41 | 42 | 43 | 44 | 45 | 46 | 47 | 48 | 49 | 50 | 51 | 52 | 53 | 54 | 55 | 56 | 57 | 58 | 59 | 60 |
|---|---|---|---|---|---|---|---|---|---|---|---|---|---|---|---|---|---|---|---|---|
| 41 | 0,167555 | | | | | | | | | | | | | | | | | | | |
| 42 | | 0,157593 | | | | | | | | | | | | | | | | | | |
| 43 | 0,088079 | | 0,172958 | | | | | | | | | | | | | | | | | |
| 44 | | 0,08547 | | 0,15445 | | | | | | | | | | | | | | | | |
| 45 | 0,041636 | | 0,098563 | | 0,172824 | | | | | | | | | | | | | | | |
| 46 | | 0,044568 | | 0,085938 | | 0,151442 | | | | | | | | | | | | | | |
| 47 | 0,088827 | | 0,043214 | | 0,102436 | | 0,172315 | | | | | | | | | | | | | |
| 48 | | 0,091146 | | 0,035897 | | 0,086124 | | 0,148559 | | | | | | | | | | | | |
| 49 | 0,163161 | | 0,071383 | | 0,051692 | | 0,103152 | | 0,167639 | | | | | | | | | | | |
| 50 | | 0,16307 | | 0,083933 | | 0,04 | | 0,083885 | | 0,15587 | | | | | | | | | | |
| 51 | | | 0,143445 | | 0,061141 | | 0,049858 | | 0,101641 | | 0,171456 | | | | | | | | | |
| 52 | | | 0,152993 | | | 0,077605 | | 0,043384 | | 0,094758 | | 0,152256 | | | | | | | | |
| 53 | | | | 0,130648 | | | 0,052259 | | 0,050288 | | 0,108897 | | 0,171403 | | | | | | | |
| 54 | | | | | 0,144033 | | | 0,072017 | | 0,047809 | | 0,093633 | | 0,151833 | | | | | | |
| 55 | | | | | | 0,121151 | | | 0,04846 | | 0,057776 | | 0,10986 | | 0,166597 | | | | | |
| 56 | | | | | | | 0,137667 | | | 0,055449 | | 0,049908 | | 0,093913 | | 0,153722 | | | | |
| 57 | | | | | | | | 0,114608 | | | 0,043051 | | 0,060687 | | 0,107518 | | 0,168684 | | | |
| 58 | | | | | | | | | 0,11943 | | | 0,051693 | | 0,048193 | | 0,098387 | | 0,151286 | | |
| 59 | | | | | | | | 0,173608 | | | 0,099956 | | | 0,041794 | | 0,059975 | | 0,112281 | | 0,165386 |
| 60 | | | | | | | | | | 0,176667 | | 0,113333 | | | 0,045 | | 0,054313 | | 0,098039 | 0,151344 |
| 61 | | | | | | | | | | | 0,158951 | | 0,091766 | | 0,045436 | | 0,06499 | | 0,109609 | |
| 62 | | | | | | | | | | | | 0,170047 | | 0,106084 | | 0,048062 | | 0,053812 | | 0,098731 |
| 63 | | | | | | | | | | | | | 0,148944 | | 0,087524 | | 0,040956 | | 0,063838 | |
| 64 | | | | | | | | | | | | | | 0,161054 | | 0,095168 | | 0,049347 | | 0,055944 |
| 65 | | | | | | | | | | | | | | | 0,142806 | | 0,076452 | | 0,042659 | |
| 66 | | | | | | | | | | | | | | | | 0,14876 | | 0,089532 | | 0,04235 |
| 67 | | | | | | | | | | | | | | | | | 0,13157 | | 0,073245 | |
| 68 | | | | | | | | | | | | | | | | | | 0,142672 | | 0,083009 |
| 69 | | | | | | | | | | | | | | | | | | | 0,126558 | |
| 70 | | | | | | | | | | | | | | | | | | | | 0,134639 |
| 71 | | | | | | | | | | | | | | | | | | | 0,175068 | |
| 72 | | | | | | | | | | | | | | | | | | | | 0,181713 |

Table 3.

| 0 | 61 | 62 | 63 | 64 | 65 | 66 | 67 | 68 | 69 | 70 | 71 | 72 | 73 | 74 | 75 | 76 | 77 | 78 |
|---|---|---|---|---|---|---|---|---|---|---|---|---|---|---|---|---|---|---|
| 61 | 0,167746 | | | | | | | | | | | | | | | | | |
| 62 | | 0,152116 | | | | | | | | | | | | | | | | |
| 63 | 0,114403 | | 0,1664 | | | | | | | | | | | | | | | |
| 64 | | 0,101583 | | 0,153656 | | | | | | | | | | | | | | |
| 65 | 0,070342 | | 0,115053 | | 0,162994 | | | | | | | | | | | | | |
| 66 | | 0,060209 | | 0,105068 | | 0,154831 | | | | | | | | | | | | |
| 67 | 0,042596 | | 0,071134 | | 0,11215 | | 0,162454 | | | | | | | | | | | |
| 68 | | 0,044814 | | 0,063804 | | 0,106853 | | 0,152747 | | | | | | | | | | |
| 69 | 0,06264 | | 0,046236 | | 0,069524 | | 0,113475 | | 0,164486 | | | | | | | | | |
| 70 | | 0,074663 | | 0,041013 | | 0,066897 | | 0,10636 | | 0,15161 | | | | | | | | |
| 71 | 0,1147 | | 0,056746 | | 0,047156 | | 0,072214 | | 0,117365 | | 0,165281 | | | | | | | |
| 72 | | 0,125 | | 0,065972 | | 0,046485 | | 0,067538 | | 0,106736 | | 0,148769 | | | | | | |
| 73 | 0,163195 | | 0,108417 | | 0,054779 | | 0,044531 | | 0,076413 | | 0,118802 | | 0,16488 | | | | | |
| 74 | | 0,171961 | | 0,116101 | | 0,059146 | | 0,046088 | | 0,067971 | | 0,104228 | | 0,151487 | | | | |
| 75 | | | 0,155634 | | 0,104837 | | 0,048636 | | 0,051296 | | 0,079139 | | 0,119971 | | 0,162859 | | | |
| 76 | | | | 0,161994 | | 0,107996 | | 0,055036 | | 0,04467 | | 0,066471 | | 0,108534 | | 0,153034 | | |
| 77 | | | | | 0,150731 | | 0,097411 | | 0,047996 | | 0,050533 | | 0,081497 | | 0,119322 | | 0,164489 | |
| 78 | | | | | | 0,15286 | | 0,102564 | | 0,050296 | | 0,045279 | | 0,071956 | | 0,111501 | | 0,153589 |
| 79 | | | | | | | 0,143101 | | 0,088586 | | 0,045488 | | 0,050692 | | 0,08099 | | 0,121549 | |
| 80 | | | | | | | | 0,147142 | | 0,09747 | | 0,048735 | | 0,04649 | | 0,075109 | | 0,1125 |
| 81 | | | | | | | | | 0,133303 | | 0,081463 | | 0,043189 | | 0,049096 | | 0,084381 | |
| 82 | | | | | | | | | | 0,140946 | | 0,094558 | | 0,047111 | | 0,051769 | | 0,077114 |
| 83 | | | | | | | | | | | 0,174565 | | 0,125193 | | 0,074939 | | 0,042823 | | 0,053288 |
| 84 | | | | | | | | | | | | 0,181122 | | 0,136905 | | 0,085034 | | 0,046453 | 0,050903 |
| 85 | | | | | | | | | | | | | 0,166352 | | 0,118463 | | 0,071414 | | 0,04235 |
| 86 | | | | | | | | | | | | | | 0,176805 | | 0,127332 | | 0,077859 | 0,044248 |
| 87 | | | | | | | | | | | | | | | 0,158749 | | 0,113851 | | 0,064141 |
| 88 | | | | | | | | | | | | | | | | 0,166538 | | 0,119287 | 0,072037 |
| 89 | | | | | | | | | | | | | | | | | 0,153227 | | 0,105727 |
| 90 | | | | | | | | | | | | | | | | | | 0,157778 | 0,112593 |
| 91 | | | | | | | | | | | | | | | | | | 0,145028 | |
| 92 | | | | | | | | | | | | | | | | | | | 0,150957 |

Table 4.

| | 79 | 80 | 81 | 82 | 83 | 84 | 85 | 86 | 87 | 88 | 89 | 90 | 91 | 92 | 93 |
|---|---|---|---|---|---|---|---|---|---|---|---|---|---|---|---|
| 79 | 0,162283 | | | | | | | | | | | | | | |
| 80 | | 0,150478 | | | | | | | | | | | | | |
| 81 | 0,120595 | | 0,164055 | | | | | | | | | | | | |
| 82 | | 0,110493 | | 0,151401 | | | | | | | | | | | |
| 83 | 0,084399 | | 0,123719 | | 0,162452 | | | | | | | | | | |
| 84 | | 0,075949 | | 0,112623 | | 0,151515 | | | | | | | | | |
| 85 | 0,054739 | | 0,087811 | | 0,122487 | | 0,161943 | | | | | | | | |
| 86 | | 0,050901 | | 0,078254 | | 0,113112 | | 0,150826 | | | | | | | |
| 87 | 0,04505 | | 0,057372 | | 0,087484 | | 0,123131 | | 0,162498 | | | | | | |
| 88 | | 0,042275 | | 0,055101 | | 0,079627 | | 0,11348 | | 0,152331 | | | | | |
| 89 | 0,060525 | | 0,042946 | | 0,057696 | | 0,089045 | | 0,124811 | | 0,160746 | | | | |
| 90 | | 0,06963 | | 0,047515 | | 0,054054 | | 0,080822 | | 0,116066 | | 0,153074 | | | |
| 91 | 0,101813 | | 0,05347 | | 0,045283 | | 0,059223 | | 0,090955 | | 0,123374 | | 0,161008 | | |
| 92 | | 0,109851 | | 0,063785 | | 0,042687 | | 0,053705 | | 0,083606 | | 0,117168 | | 0,153569 | |
| 93 | 0,140228 | | 0,093953 | | 0,050482 | | 0,045322 | | 0,061825 | | 0,090312 | | 0,124674 | | 0,159175 |
| 94 | | 0,147318 | | 0,103191 | | 0,059063 | | 0,043478 | | 0,055808 | | 0,085518 | | 0,118634 | |
| 95 | 0,176046 | | 0,131698 | | 0,090039 | | 0,048314 | | 0,04658 | | 0,061774 | | 0,092454 | | 0,12373 |
| 96 | | 0,182292 | | 0,139974 | | 0,097656 | | 0,054688 | | 0,041613 | | 0,058313 | | 0,087761 | |
| 97 | | | 0,167499 | | 0,127557 | | 0,085038 | | 0,048131 | | 0,048359 | | 0,063937 | | 0,091643 |
| 98 | | | | 0,174891 | | 0,134291 | | 0,093067 | | 0,04872 | | 0,045872 | | 0,060534 | |
| 99 | | | | | 0,162622 | | 0,121812 | | 0,079147 | | 0,047993 | | 0,046953 | | 0,063696 |
| 100 | | | | | | 0,168566 | | 0,128974 | | 0,086383 | | 0,048474 | | 0,046866 | |
| 101 | | | | | | | 0,156245 | | 0,115253 | | 0,076043 | | 0,044884 | | 0,047942 |
| 102 | | | | | | | | 0,16263 | | 0,121684 | | 0,080738 | | 0,047646 | |
| 103 | | | | | | | | | 0,149607 | | 0,11192 | | 0,070807 | | 0,044804 |
| 104 | | | | | | | | | | 0,155297 | | 0,115918 | | 0,07543 | |
| 105 | | | | | | | | | | | 0,145584 | | 0,106029 | | 0,068123 |
| 106 | | | | | | | | | | | 0,186866 | | 0,148959 | | 0,109984 |
| 107 | | | | | | | | | | | | 0,177225 | | 0,139135 | 0,102632 |
| 108 | | | | | | | | | | | | | 0,180041 | | 0,14249 |
| 109 | | | | | | | | | | | | | | 0,170722 | 0,135559 |
| 110 | | | | | | | | | | | | | | 0,173525 | |
| 111 | | | | | | | | | | | | | | | 0,166564 |

Table 5.

## 8. Proximity graph

In this section we describe the proximity of disc-like octagons and discs. Blue squares denote disc-like octagons and the number written inside the square after the letter "o" denotes the diameter of the octagon. Similarly, red discs denote discs and the number after the letter "c" denotes the diameter of the disc.

An arrow indicates the nearest shape for the source of the arrow and the label is truncated Jaccard's distance between these shapes.

The graph is divided into two components: the component of shapes with odd diameters (the odd component) and the component of shapes with even diameters (the even component). Each of these components are presented in two parts so that the material fits on one page.

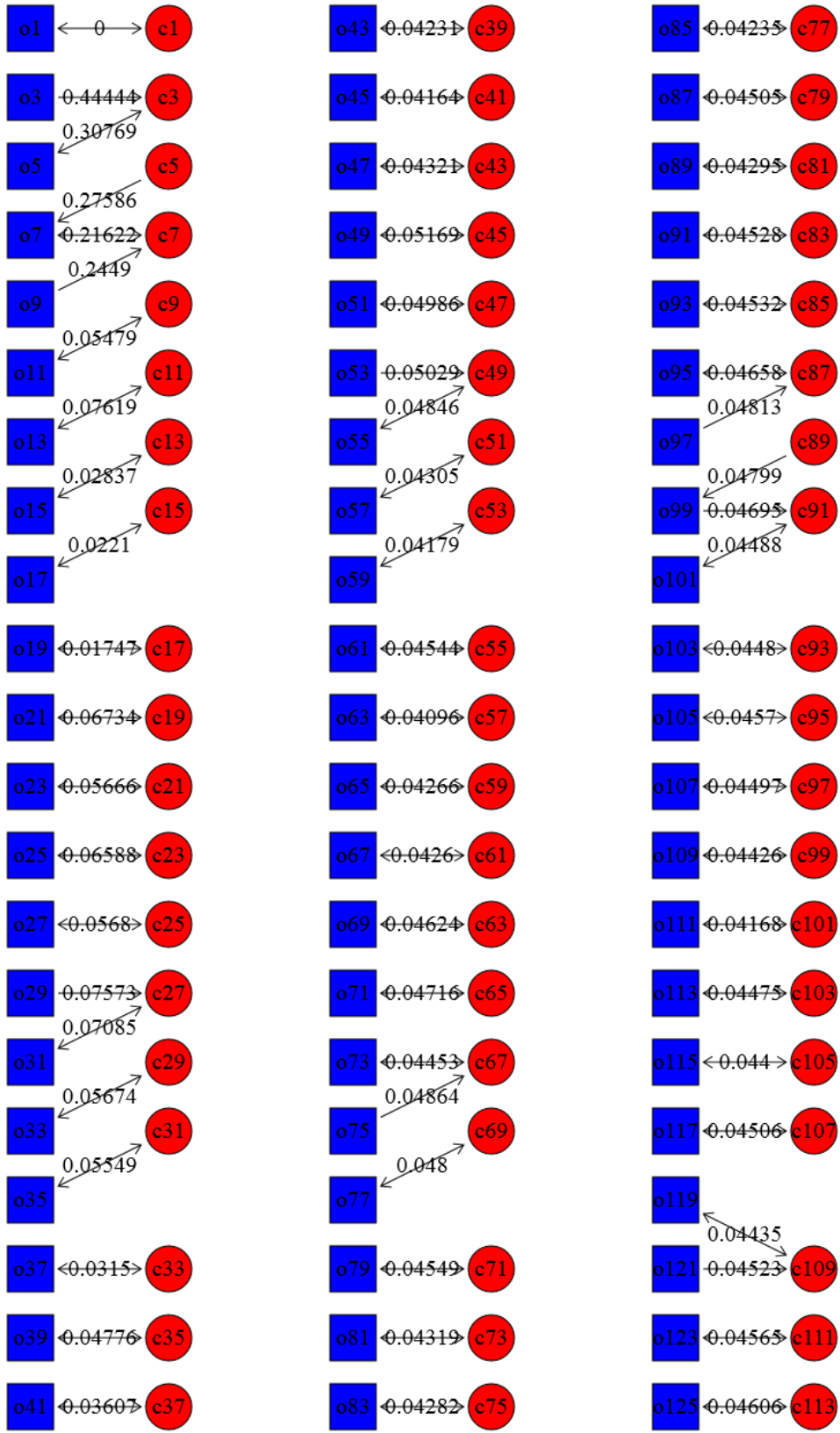

Proximity graph. The odd component. Part 1.

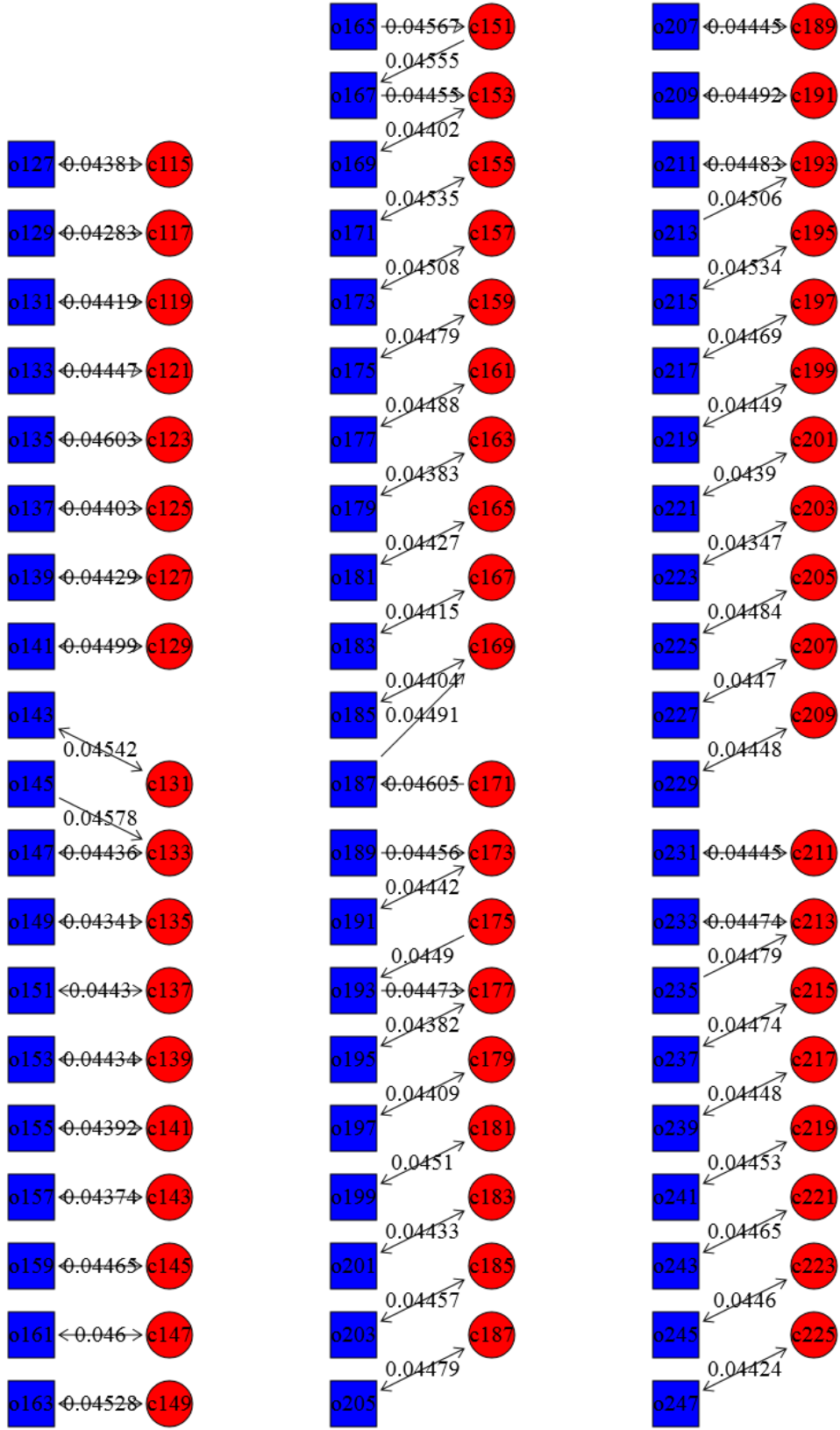

Proximity graph. The odd component. Part 2.

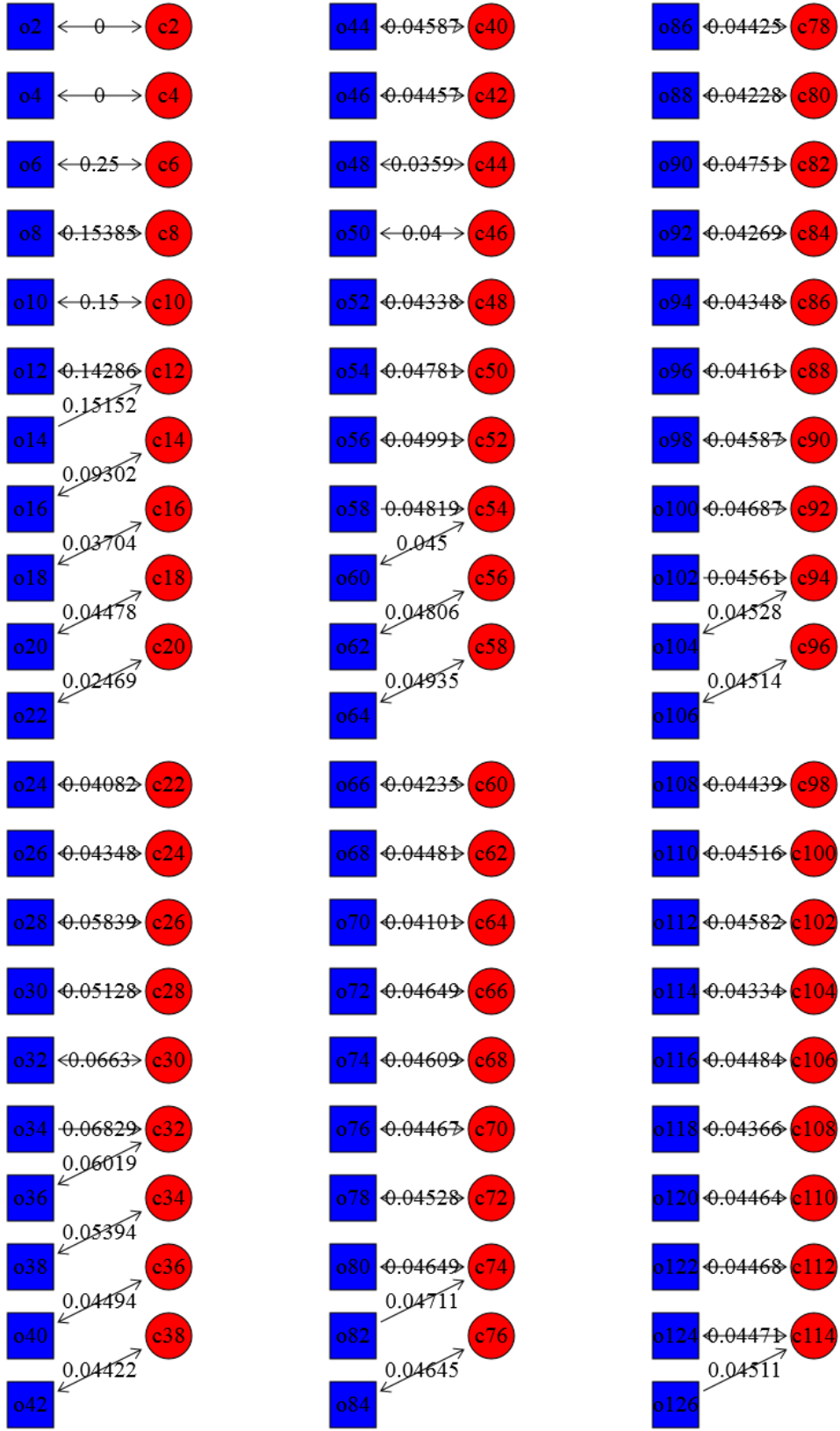

Proximity graph. The even component. Part 1.

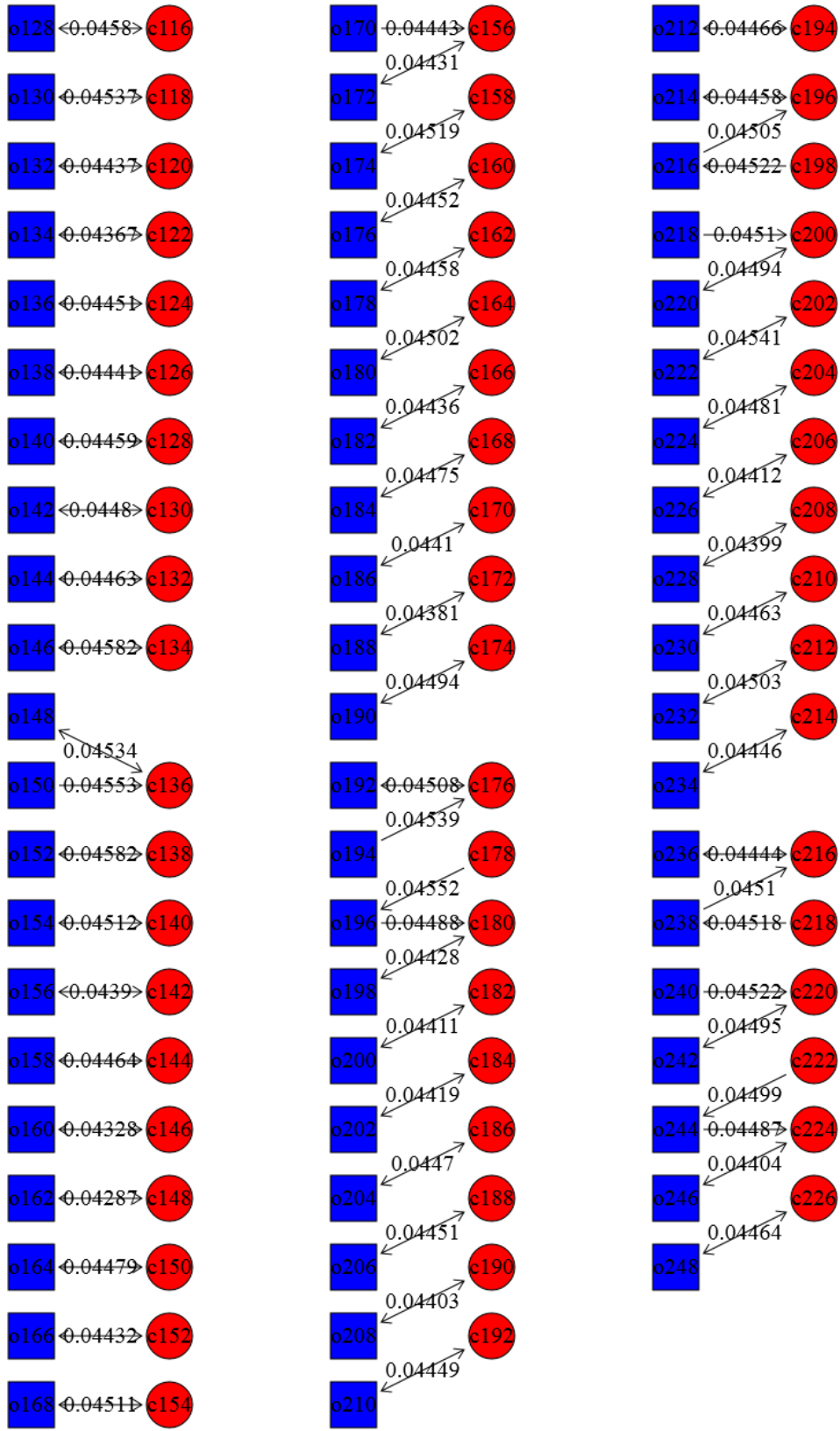

Proximity graph. The even component. Part 2.